\documentclass[12pt, onecolumn, draftclsnofoot, journal]{IEEEtran}

\usepackage{amsfonts}

\usepackage{textcomp}
\usepackage{graphicx}
\usepackage{algorithm,algorithmic,amsbsy,amsmath,amssymb,epsfig,subfig,bbm,mathrsfs,mathdots,arydshln,arydshln,multirow,verbatim,booktabs,setspace,url,stfloats,cite,nomencl,epstopdf, comment}

\usepackage{multicol}
\usepackage{braket}

\usepackage{xcolor}
\def\BibTeX{{\rm B\kern-.05em{\sc i\kern-.025em b}\kern-.08em
    T\kern-.1667em\lower.7ex\hbox{E}\kern-.125emX}}

\newcommand{\beq}{\begin{equation}}
\newcommand{\eeq}{\end{equation}}
\newcommand{\bitm}{\begin{itemize}}
\newcommand{\ba}{\begin{array}}
\newcommand{\ea}{\end{array}}
\newcommand{\eitm}{\end{itemize}}
\newcommand{\beqn}{\begin{eqnarray}}
\newcommand{\eeqn}{\end{eqnarray}}
\newcommand{\beqno}{\begin{eqnarray*}}
\newcommand{\eeqno}{\end{eqnarray*}}
\newcommand{\bma}{\begin{displaymath}}
\newcommand{\ema}{\end{displaymath}}
\newcommand{\bnu}{\begin{enumerate}}
\newcommand{\enu}{\end{enumerate}}
\newcommand{\bce}{\begin{center}}
\newcommand{\ece}{\end{center}}
\newcommand{\btb}{\begin{tabular}}
\newcommand{\etb}{\end{tabular}}

\begin{document}

\title{Elastic Entangled Pair and Qubit Resource Management in Quantum Cloud Computing}
\author{\IEEEauthorblockN{Rakpong Kaewpuang, 
Minrui Xu, 
Dinh Thai Hoang, 
Dusit Niyato,
Han Yu, 
Ruidong Li,
Zehui Xiong, and Jiawen Kang
}}
\maketitle

\begin{abstract}
Quantum cloud computing (QCC) offers a promising approach to efficiently provide quantum computing resources, such as quantum computers, to perform resource-intensive tasks. Like traditional cloud computing platforms, QCC providers can offer both reservation and on-demand plans for quantum resource provisioning to satisfy users' requirements. However, the fluctuations in user demand and quantum circuit requirements are challenging for efficient resource provisioning. Furthermore, in distributed QCC, entanglement routing is a critical component of quantum networks that enables remote entanglement communication between users and QCC providers. Further, maintaining entanglement fidelity in quantum networks is challenging due to the requirement for high-quality entanglement routing, especially when accessing the providers over long distances. To address these challenges, we propose a resource allocation model to provision quantum computing and networking resources. In particular, entangled pairs, entanglement routing, qubit resources, and circuits' waiting time are jointly optimized to achieve minimum total costs. We formulate the proposed model based on the two-stage stochastic programming, which takes into account the uncertainties of fidelity and qubit requirements, and quantum circuits' waiting time. Furthermore, we apply the Benders decomposition algorithm to divide the proposed model into sub-models to be solved simultaneously. Experimental results demonstrate that our model can achieve the optimal total costs and reduce total costs at most 49.43\% in comparison to the baseline model.       
\end{abstract}


\section{Introduction}
\label{sec:introduction}

Quantum cloud computing (QCC)~\cite{ibm-quantum-computing, quantum-ai-google2022, amazon-barket2022, azure-quantum2022} has the capability to address complex simulation and optimization challenges in communication and network systems at a large scale. By utilizing quantum bits (qubits) and employing techniques, such as superposition, entanglement, and interference, QCC has the great potential to surpass the classical cloud computing~\cite{s-chaisiri-optimization2011,s-chaisiri-ovmp2009} and the existing supercomputers by accelerating computations and lowering energy consumption. The emergence of Noisy Intermediate-Scale Quantum (NISQ) computing has spurred AWS~\cite{amazon-barket2022}, IBM~\cite{ibm-quantum-computing}, and Azure~\cite{azure-quantum2022}, offering QCC that is transforming the fields of finance, machine learning, and security. However, with the current technologies, quantum resources, such as qubits, are limited and costly in QCC as opposed to traditional cloud computing. The efficacy of quantum computing is impacted not just by the quantity of qubits, but also by the depth of the quantum circuit and the level of noise presenting at various points within the circuit. The scale, quality, and speed of QCC are all critical factors that determine the size and complexity of quantum computing tasks that can be effectively addressed. In addition, these computing tasks can be considered as random input for QCC.   

Similar to the definition in classical cloud computing, QCC operators match users of quantum cloud applications with quantum computer providers in the cloud. In QCC, a user can request the necessary quantum computing resources from a quantum cloud service provider, which is similar to traditional cloud computing. During execution, the user can specify the amount of resources required in terms of qubits and quantum circuits to the provider, depending on the complexity of the computing task. The provider can offer users two resource provisioning plans, namely reservation and on-demand plans. For the reservation plans, user reserves the required quantum computing resources from the operator based on the expected task difficulty and waiting time. Due to the uncertainties of qubit requirements and minimum waiting time for quantum circuits, the user can also purchase additional quantum computing resources from the operator for the execution of the computing tasks.

Recently, as a promising approach to support QCC and distributed QCC, quantum networks have been created to facilitate groundbreaking applications in materials science, drug discovery, and cryptography~\cite{bennett-quantum-cryptography1984, c-li-effective-routing2021, y-cao-hybrid-trusted-untrusted2021} that go beyond traditional networks. Quantum networks connect quantum nodes through optical fiber links or free space~\cite{y-cao-hybrid-trusted-untrusted2021}, where the nodes generate and store quantum information, and also transmit and receive it between each other~\cite{j-li-fidelity-guaranteed-entanglement2022, s-shi-concurrent-entanglement2020}. However, prior to information exchange, it is necessary for two quantum nodes to establish an entangled connection between them. This connection allows for the transmission of quantum information, encoded as qubits. Therefore, the quantum source node can transmit information to the quantum destination node using entangled pairs. When the source node and the destination node are distant from each other, remote entanglement connections are established according to the assigned routing. Intermediate quantum nodes, known as quantum repeaters, connect source and destination nodes using entanglement swapping, which involves joint Bell state measurements, to create a remote entanglement connection~\cite{c-li-effective-routing2021}. Therefore, a critical challenge for constructing quantum networks at a large scale is the efficient utilization of entangled pairs and the identification of optimal routing strategies for managing massive entanglement connections.

Meanwhile, maintaining Entanglement fidelity is crucial to ensure high-quality remote entanglement connections, as the noise in the system~\cite{j-li-fidelity-guaranteed-entanglement2022} may prevent quantum repeaters from producing entangled pairs with the desired fidelity. Low-fidelity entangled pairs can adversely impact the quality of services offered by quantum applications~\cite{a-s-cacciapuoti-quantum-internet2020}. For example, when the fidelity of entangled pairs falls below the quantum bit error rate (BER) in quantum cryptographic protocols, it can lead to the degradation of the security of key distribution~\cite{q-jia-an-improved2021}. Fortunately, the entanglement purification techniques~\cite{a-s-cacciapuoti-when-entanglement2020, r-van-meter-system-design2009, x-m-long-distance-2021} can improve the fidelity of entangled pairs by using additional entangled pairs. These techniques utilize multiple entangled pairs to combine them in various ways to increase the fidelity of the final purified entangled pair, such as entanglement distillation, quantum error correction, and decoherence-free subspaces. However, determining the optimal number of additional entangled pairs required by the entanglement purification technique to meet the uncertain requirements of fidelity values needed by quantum applications is challenging and has been overlooked in the literature.

To overcome the challenges discussed above, in this paper, we propose an entangled pair and qubit resource management model in QCC. We focus on entangled pair resource allocation and fidelity-guaranteed entanglement routing in quantum networks, together with qubit resource allocation for quantum applications on quantum computers of the QCC providers. Specifically, we formulate the two-stage stochastic programming model to determine the optimal number of entangled pairs and the optimal number of qubits that can fulfill all requests from multiple quantum source nodes (i.e., users) and quantum destination nodes (i.e., providers). In the optimization problem, the uncertainties of fidelity requirements, the number of qubits, and the waiting time for quantum applications are taken into consideration. In addition, we apply the Benders decomposition algorithm to reduce both the complexity and execution time of the problem. The goal of the proposed model is to make optimal decisions for quantum applications in minimizing the total costs regarding entangled pairs, qubits, and quantum applications' waiting time. The main contribution of this paper can be summarized as follows:
\begin{itemize}
    \item We introduce an innovative model of a joint entangled pair and qubit resource allocation, and entanglement routing with a fidelity guarantee under uncertainties related to fidelity requirements, qubit requirements, and quantum applications' waiting time in QCC. In addition, we introduce the dynamic entanglement purification algorithm to enhance the fidelity value at a link between two quantum nodes.
    \item We formulate the two-stage stochastic programming (SP) model to determine the optimal allocation of entangled pairs and qubit resources, as well as the fidelity-guaranteed entanglement routing and minimum waiting time of quantum applications in QCC. In the proposed model, both the entangled pair and qubit resource allocation and the fidelity-guaranteed entanglement routing are calculated at the first stage using statistical information and then refined in the second stage with actual realization.
    \item We apply the Benders decomposition algorithm to divide the proposed model into smaller models which can be solved concurrently.  
    \item To assess the effectiveness of our proposed model, we conduct extensive experiments using real-world network topology. We demonstrate the superiority of our proposed model by conducting experiments on the circuit demands of quantum Fourier transform (QFT) within practical quantum computing programming environments. In addition, we compare the outcomes of our proposed model to those of benchmark models to demonstrate its superior performance.
\end{itemize}
The remainder of this paper is organized as follows. Section~\ref{sec:related-works} reviews related work. Section~\ref{sec:system-model} describes the system model, network model, and the case study of quantum Fourier transform. Section~\ref{sec:optimization-formulation} presents the proposed stochastic programming model in QCC. Section~\ref{sec:benders-decomposition} further elaborates the Benders decomposition algorithm. Section~\ref{sec:performance-evaluation} shows the performance evaluation results. Section~\ref{sec:conclusion} concludes the paper.

\section{Related Work}
\label{sec:related-works}

\subsection{Quantum Networks}

In~\cite{j-li-fidelity-guaranteed-entanglement2022}, the authors proposed two algorithms of fidelity-guaranteed entanglement routing for quantum networks to ensure the fidelity of the entangled state between source-destination (S-D) pairs. The first algorithm (called Q-PATH) was proposed to achieve the optimal routing path and the minimum cost of the entangled pair while meeting the requirement of a single S-D pair. To reduce the computational complexity of Q-PATH, the second algorithm (called Q-LEAP) was introduced to obtain a satisfactory routing path with minimum fidelity degradation. In addition, the entanglement routing approach based on the greedy algorithm was proposed for S-D pairs to minimize the routing path and the number of entangled pairs. Similarly, in~\cite{c-li-effective-routing2021}, the authors introduced a general routing scheme for generating entanglements on a quantum lattice network with limited quantum resources, specifically the quantity of quantum memories in each node. The objective of this scheme was to allocate quantum resources effectively to meet the requests on entanglement generations and the desired fidelity thresholds of entanglements.

In~\cite{k-chakraborty-entanglement-dist2020}, the authors proposed an efficient linear programming model to maximize the entanglement distribution rate between multiple S-D pairs in a quantum network, while maintaining the desired end-to-end fidelity. Although this problem was similar to the one addressed in~\cite{j-li-fidelity-guaranteed-entanglement2022}, the purification process was not considered in~\cite{k-chakraborty-entanglement-dist2020}. In~\cite{y-zhao-redundant-entanglement2021}, the authors proposed a novel redundant entanglement provisioning and selection (REPS) scheme to maximize throughput for multiple S-D pairs in a multi-hop quantum network. REPS was designed to support multiple entanglement routing and cope with entanglement generation failures. In~\cite{l-gyongyosi-adaptive-routing2019}, the authors presented a dynamic adaptive routing scheme to handle the potential failure of quantum memories in quantum nodes in the quantum network. The primary goal of this approach was to determine the shortest node-disjoint replacement paths in the network of the quantum Internet and minimize the lost entangled contacts. However, in the event of quantum memory failures, replacement paths were implemented to temporarily establish entangled connections and ensure uninterrupted network transmission. In~\cite{m-caleffi-optimal-routing2017}, the author introduced the joint routing protocol design and route metric table for quantum networks with the objective of identifying the path with the highest possible end-to-end entanglement rate between the S-D pairs. The authors of~\cite{f-hahn-quantum-network2019} applied graph theoretic tools, specifically graph states, to minimize the quantity of required measurements and develop a routing approach for quantum communication between the S-D nodes.

\subsection{Quantum Cloud Computing}

In the literature, several studies have proposed resource management schemes for quantum computing. For example, in \cite{n-ngoenriang-optimal-stochastic2022}, a two-stage stochastic programming approach was used to obtain the minimum deployment cost for quantum resources while accounting for uncertainties in computational works, quantum computer availability and computing power, and fidelity of entangled qubits. Another study, \cite{g-s-ravi-quantum-computing-in-cloud2021}, focused on analyzing the computation time of works and resource utilization in IBM quantum cloud systems. They analyzed waiting times, computation time of quantum machines and circuits, and machine utilization. In \cite{g-s-ravi-adaptive-job2021}, the authors introduced an optimized adaptive job scheduling approach for IBM quantum cloud systems, which reduced waiting times and improved fidelity. Quantum resource scheme for distributed quantum computing was introduced in \cite{c-cicconetti-resource-allocation-quantum2022}. The objective of the scheme was to compute traffic flows for all paths of applications in advance, and then allocate resources (e.g., gross rates) to applications by using the round-robin strategy. In \cite{r-kaewpuang-stochastic-qubit2023}, the authors proposed a stochastic quantum resource management scheme for QCC systems, where the allocation of qubit resources and minimum waiting time for quantum circuits were jointly calculated to minimize the total cost of quantum circuits while accounting for uncertainties in the required qubits and minimum waiting time.

However, the existing work fails to address the challenge of simultaneously optimizing the allocation of entangled pairs, the routing of entanglement with guaranteed fidelity, and the allocation of qubit resources for quantum circuits in QCC. Moreover, current approaches overlook the uncertainties associated with fidelity requirements, the number of required qubits, and the waiting time for quantum circuits, which can significantly impact the overall cost of quantum circuits in QCC.

\section{System Model}
\label{sec:system-model}
We first introduce the QCC environment, the quantum network, and the relationship among quantum components. Then, we describe the components and entanglement operations in a quantum network. Finally, we present the QFT circuit, which is the case study for the computing application.

\begin{figure}[htb]
\centering
\captionsetup{justification=centering}
$\begin{array}{c} \epsfxsize=3.8 in \epsffile{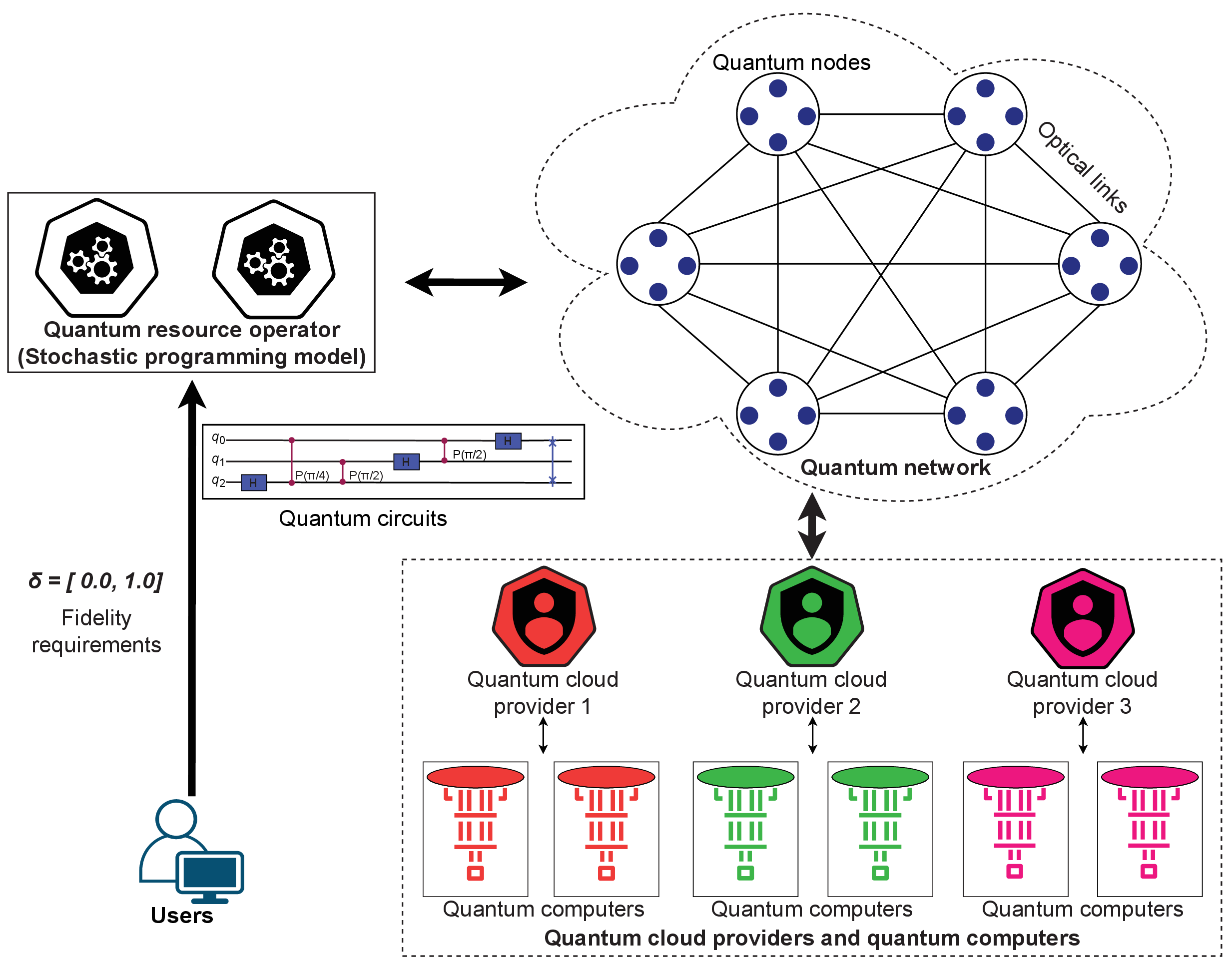} \\
\end{array}$
\caption{The QCC environment.} 
\label{fig:system-model}
\end{figure}
We consider the QCC environment illustrated in Fig.~\ref{fig:system-model}. The system model consists of users, QCC providers, quantum computers, a quantum network, and a quantum resource operator. Users possess quantum circuits that they want to execute on the quantum computers provided by QCC providers. Users request fidelity requirements (i.e., quality of entanglement qubits), a number of qubits for executing quantum circuits, and a waiting time for quantum circuits to be completed. QCC providers provide quantum computing resources, such as qubits in quantum computers, to users. 

We consider fidelity requirements, the number of qubits, and the waiting time for quantum circuits as uncertain demands. The fidelity requirement, the number of qubits, and the waiting time for quantum circuit $c$ are denoted by $\tilde{\theta_{c}}$, $\tilde{\beta_{c}}$ and $\tilde{\alpha_{c}}$, respectively. QCC providers offer both reservation and on-demand plans to users for quantum computing resource provision, which are similar to the pricing plans used in conventional cloud computing~\cite{s-chaisiri-optimization2011,s-chaisiri-ovmp2009}. The cost of the on-demand plan is practically higher than that of the reservation plan. Three phases~\cite{s-chaisiri-optimization2011,s-chaisiri-ovmp2009} are introduced to provision computing resources: reservation, utilization, and on-demand. During the reservation phase, computing resources are allocated to users without information about their specific requirements, and then the reserved computing resources are used during the utilization phase. However, if the reserved computing resources are insufficient, the computing resources during the on-demand phase are allocated for satisfying the remaining requirements. From users' perspectives, quantum circuits are successfully executed as fast as possible. Therefore, for the specific waiting time, the QCC providers will incur a penalty cost of over-waiting time if the circuits cannot be completed in the specified waiting time. 

For the quantum network, the entangled pair resource allocation and entanglement routing with fidelity guarantee are crucial to ensure the fidelity requirements of users for quantum applications, specifically quantum circuits. In Fig.~\ref{fig:system-model}, quantum nodes are interconnected in the network through optical links and have the capability to create, exchange, store, and process quantum information~\cite{j-li-fidelity-guaranteed-entanglement2022,s-shi-concurrent-entanglement2020}, as depicted in Figs.~\ref{fig:one-teleportation-three-components}(a) and (b). Figures~\ref{fig:one-teleportation-three-components}(a) and (b) illustrate the process of transmitting information from a quantum source node to a quantum destination node in the quantum network. To create the entanglement connection between distant quantum source and destination nodes, entangled pairs of intermediate quantum nodes are generated. Quantum repeaters, which are intermediate quantum nodes, perform entanglement swapping to create long-distance entanglement connections between source and destination nodes. Taking Fig.~\ref{fig:one-teleportation-three-components}(b) as an example, when source 2 transmits one qubit (information) to destination 2, quantum repeater entangles with both nodes and performs entanglement swapping to establish a connection between them to achieve such transmission. The quantum network is represented by a graph~{\cite{f-hahn-quantum-network2019, m-caleffi-optimal-routing2017,l-gyongyosi-adaptive-routing2019}, consisting of quantum nodes and edges. Sets of quantum nodes and edges (links) are defined as $\mathcal{N}$ and $\mathcal{L}$, respectively. Each quantum node has limited quantum memories and a restricted number of entangled pairs. To ensure the fidelity of information transmission, entanglement purification will function on the specific nodes to meet the fidelity requirement and the fidelity threshold. The entanglement purification operation utilizes the entangled pairs to improve the fidelity values on edges. The fidelity value on the same edge is the same, while the fidelity value on different edges is likely different~\cite{c-li-effective-routing2021}.

\begin{figure*}[htb]
 \centering
 \captionsetup{justification=centering}
 \subfloat{\label{fig:teleportation}\includegraphics[width=0.50\textwidth]{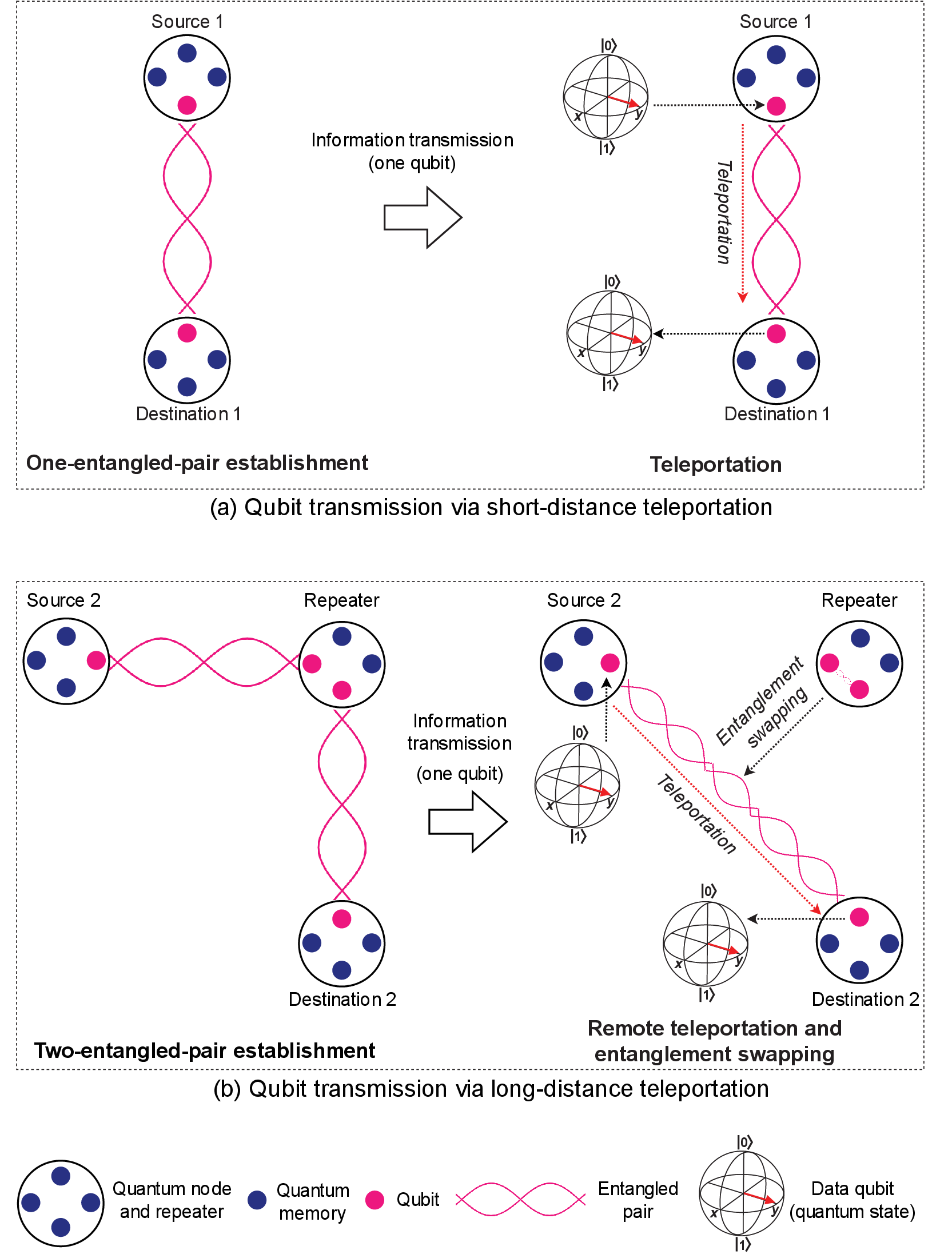}}
 \subfloat{\label{fig:three-components}\includegraphics[width=0.45\textwidth]{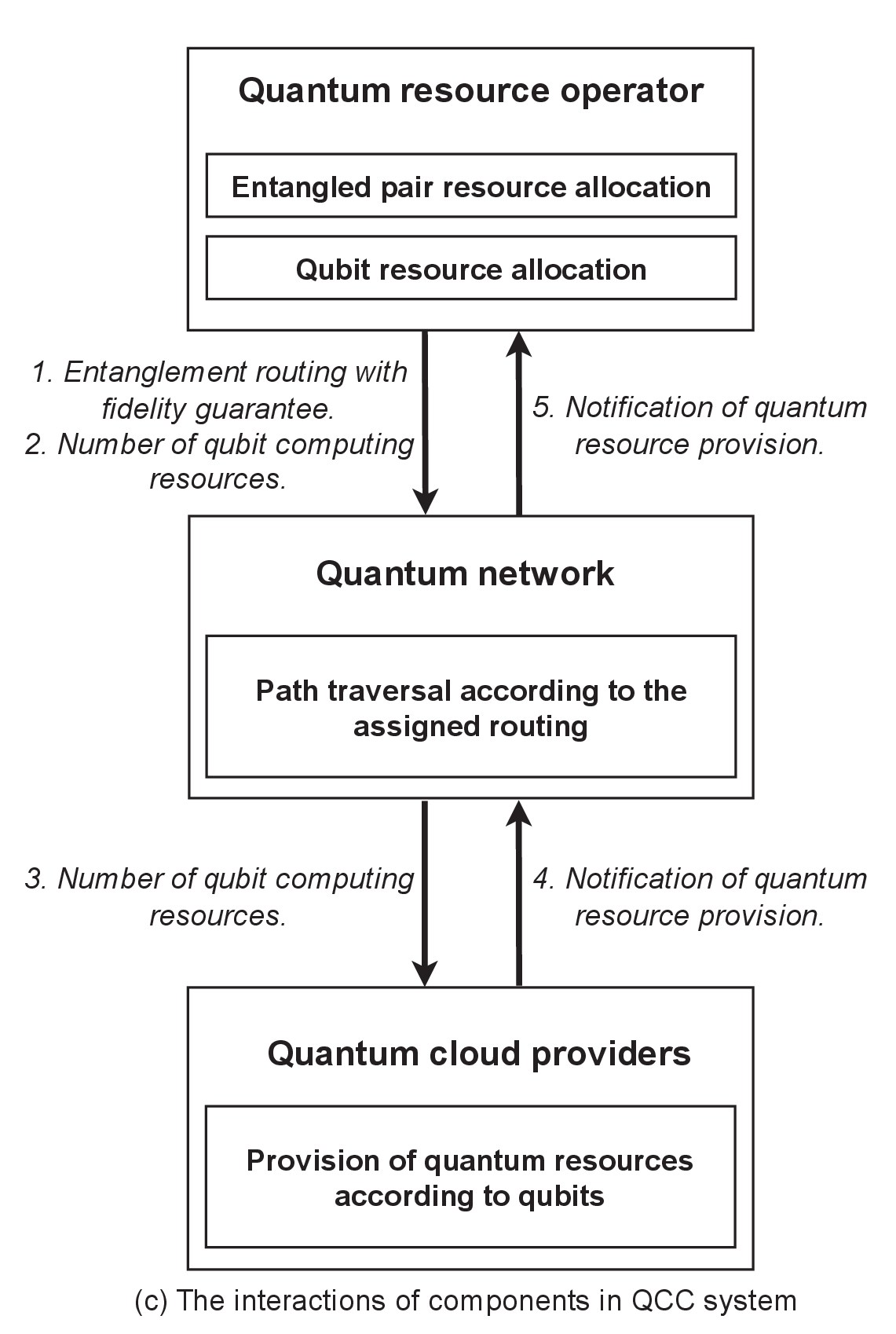}}
 \caption{(a) One qubit is sent by short-distance quantum teleportation, (b) One qubit is sent via entanglement swapping by long-distance quantum teleportation, and (c) The interactions of the quantum resource operator, quantum network, and QCC providers.}
 \label{fig:one-teleportation-three-components}
\end{figure*} 

To achieve the lowest possible cost associated with entangled pair resources, guaranteed-fidelity entanglement routings, qubit computing resources, and the penalty cost of waiting too long while meeting uncertainties of fidelity and user requirements, the quantum resource operator is designed to provide the optimal entangled pair resources and routings in QCC. In addition, the operator efficiently provisions the quantity of qubit computing resources and the minimum waiting time for quantum circuits. The quantum resource operator is formulated using the two-stage SP.

Figure~\ref{fig:one-teleportation-three-components}(c) illustrates the sequence of interactions among the quantum resource operator, quantum network, and QCC providers. In the operator, the entangled pair resource allocation assigns the entanglement routing that meets users' fidelity requirements and provides the number of entangled pairs. The qubit resource allocation provides the quantity of quantum computing components, e.g., qubits and quantum gates, that satisfy users' qubit requirements. These quantum computing resources, e.g., qubits, are related to the quantum computing application provisioning of QCC providers. In the quantum network, quantum nodes are selected and utilized to establish connections from source nodes (i.e., users) to destination nodes (i.e., providers) with the assigned entanglement routing and qubits. For providers, the provider and its quantum computers are allocated according to the assigned qubits. If quantum computing takes longer than the expected execution time to run the quantum circuit, the provider has to pay a penalty cost for the excess waiting time. Notification is then sent to the operator through the network. If the entangled pairs and qubits are insufficient, the entangled pair and qubit resource allocation are recalculated. Otherwise, the entangled pair and qubit resource allocation stop.

\subsection{Network Model}

The components of the quantum network are quantum node, quantum source, quantum destination, quantum repeater, and quantum channel.

\subsubsection{Quantum node, quantum source, quantum destination, and quantum repeater} The quantum node performs various functions such as generating and processing quantum information, establishing quantum networks, and supporting quantum applications~\cite{s-shi-concurrent-entanglement2020}. It also contains a quantum repeater function, which involves entanglement generation, purification, and swapping~\cite{j-li-fidelity-guaranteed-entanglement2022}. The quantum node is often referred to as the quantum repeater. In quantum networks, quantum nodes have limited computing and storage capacities, and are connected via classical networks~\cite{j-li-fidelity-guaranteed-entanglement2022}. The quantum nodes are controlled by the network controller that has all information about the network, e.g., network topology and available resources. A quantum source node establishes an entanglement connection with a quantum destination node based on the requirements of the quantum application.

\subsubsection{Quantum channel} The quantum channel is established between adjacent quantum nodes to transmit qubit information via optical fibers~\cite{j-li-fidelity-guaranteed-entanglement2022, s-shi-concurrent-entanglement2020} or free space~\cite{j-g-ren-ground-to-satellite-quantum-teleportation}. Each channel shares entangled pairs of adjacent quantum nodes, and its capacity is determined by the quantity of entangled pairs generated through the entanglement generation process (e.g., nitrogen-vacancy centers~\cite{p-c-humphreys-deterministic-delivery2018}). The fidelity of the entangled pair is calculated using a deterministic equation without noise~\cite{p-c-humphreys-deterministic-delivery2018},~\cite{m-caleffi-quantum-switch2020}.

The entanglement routing process consists of three steps. Firstly, entangled pairs are generated between quantum source and destination nodes, and adjacent nodes connecting with the quantum channel. Then, the routing and entangled pair resources are allocated by the network controller in the network, respectively. Finally, the corresponding quantum nodes operate entanglement purification to enhance the fidelity of the entangled pairs and meet applications' requirements, which are instructed by the network controller. 

For multi-hop entanglement connections, entanglement swapping is used to establish remote-distance entanglement. Entanglement generation, swapping, and purification are crucial for establishing entanglement connections in quantum networks.

\subsubsection{ Entanglement generation and distribution} To create entangled pairs for two quantum nodes, a heralding station, such as nitrogen-vacancy centers in diamond~\cite{a-dahlberg-link-layer2019}, is used to perform the physical entanglement generation over optical fibers. The generated entangled pairs are distributed and stored in the memories of two quantum nodes as resources for entanglement communication and qubit transmission.

\subsubsection{Entanglement swapping} Entanglement swapping is employed to establish remote entanglement connections according to the routing when the quantum source node and quantum destination node are located far apart. By repeating the swapping operations, a multi-hop entanglement connection along the path of repeaters containing entangled pairs is established.

\begin{figure}[htb]
\centering
\captionsetup{justification=centering}
$\begin{array}{c} \epsfxsize=3.3 in \epsffile{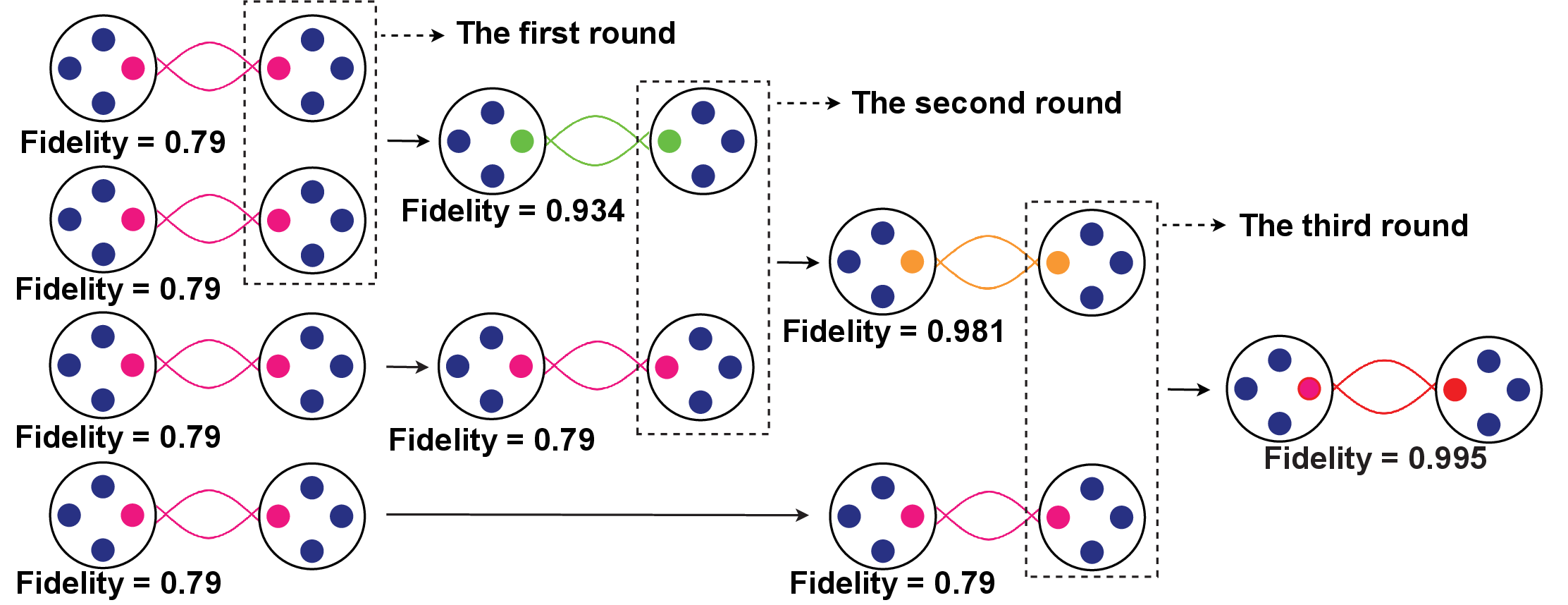} \\
\end{array}$
\caption{The example of three-round purification operations to improve a fidelity value.} 
\label{fig:purification-process}
\end{figure}

\subsubsection{Entanglement purification} Entanglement purification is applied to enhance the fidelity of a Bell pair by merging two lower-fidelity Bell pairs into a higher-fidelity Bell pair, which are implemented by using a polarizing beam splitter~\cite{x-m-long-distance-2021} or {\bf C-NOT} (controlled-NOT) gates. The entanglement purification operation~\cite{j-li-fidelity-guaranteed-entanglement2022} yields an improved fidelity  presented as follows:
\beqn
    \mathbf{F}^{\mathrm{ep}}(b_{1}, b_{2}) = \frac{b_1 b_2}{b_1 b_2 + (1-b_1)(1-b_2)}. \label{entanglement-purification-function}
\eeqn
$b_1$ and $b_2$ are the measured fidelity of two Bell pairs. In this paper, we propose an entanglement purification algorithm to dynamically calculate the operation above. In the proposed algorithm, each round of the operation applies an extra entangled pair~\cite{r-van-meter-system-design2009} to improve the pair's fidelity. For instance, by implementing three rounds of entanglement purification operations with utilization of entangled pairs, the fidelity can improve from 0.79 to 0.995 as shown in Fig.~\ref{fig:purification-process}. The entanglement purification algorithm is presented in {\bf Algorithm~\ref{algorithm-entanglement-purification}}.

\begin{algorithm}
    \caption{Entanglement purification $\mathbf{F}^{\mathrm{dep}}(\;\cdot\;)$ } \label{algorithm-entanglement-purification}
    \begin{multicols}{2}
    \begin{algorithmic}[1]
    \STATE \textbf{Input:} $y^{\mathrm{eep}}_{i,n,r,\psi}$ and $y^{\mathrm{oep}}_{i,n,r,\psi}$  \\
    \STATE \textbf{Output:} The enhanced fidelity ($ef$) \\
    \STATE  $M^{\mathrm{ep}}$ = maximum number of entangled pairs \\
    \STATE  $ef$ = 0 
        \FOR{ $pr$ = $1$ to $M^{\mathrm{ep}} -1$}   
            \IF {$pr$ == 1} 
                  \STATE $b_1$ = the first pair's fidelity
                  \STATE $b_2$ = the second pair's fidelity 
                  \STATE $ef = \mathbf{F}^{\mathrm{ep}}(b_1,b_2)$
            \ELSE 
                  \STATE $ef$ =  $ef$ of $pr - 1$
                  \STATE $b_2$ = the next pair's fidelity 
                  \STATE  $ef = \mathbf{F}^{\mathrm{ep}}(ef,b_2)$
            \ENDIF
            \STATE $pr = pr + 1$
        \ENDFOR
    \end{algorithmic}
    \end{multicols}
\end{algorithm}

\subsection{Computing Model: The Case Study of QFT Circuit}
\label{quantum-fourier-transform}
Next, we present an introduction to QFT, one of the most commonly used quantum algorithms with numerous applications in signal processing and statistical analysis. We offer a concise introduction to QFT, encompassing the necessary quantum gates and circuits needed for its implementation. Our goal is to indicate that different QFT settings demand varying numbers of qubits and computational time, both of which are essential for our proposed QCC. We mention that our proposed approach for quantum resource allocation is versatile and can be utilized in other quantum algorithms.

\subsubsection{Fundamental concept of QFT} The QFT is a quantum transformation that applies the discrete Fourier transform to the amplitudes of a wave function~\cite{quantum-fourier-transform-qiskit2022, y-s-weinstein-implementationqft2001, nielsen-quantum-computation2010}. This transformation generates a quantum state that is similar to the result of the discrete Fourier transform. Mathematically, the mapping of quantum state $\ket{U}$ to quantum state $\ket{V}$ in QFT is expressed as follows:
\beqn
        \ket{U} = \sum_{h=0}^{M-1} u_{h} \ket{h} \mapsto \ket{V} = \sum_{h'=0}^{M-1} v_{h'} \ket{h'}.
\eeqn
The amplitude $v_{h'}$ in QFT represents the discrete Fourier transform of the amplitudes $u_h$. QFT can transform a quantum state between the computational basis and the Fourier basis by utilizing quantum gates. The QFT's transformation between the states in the Fourier basis and computational basis is expressed mathematically in Eq.~(\ref{qft-eq-5}) as
\beqn
	   \ket{\tilde{u}} = \frac{1}{\sqrt{M}} \sum_{v_1 = 0}^{1} \dots \sum_{v_l = 0}^{1} \Pi_{k=1}^{l} e^{2 \pi i \frac{u v_k}{2^{k}} } \ket{v_1,v_2,\dots,v_{l}} \label{qft-eq-4} \nonumber \quad\quad\quad\quad\quad\quad\quad\quad\quad\quad\quad\quad\quad \\
                     \quad\quad\quad\quad\quad = \frac{1}{\sqrt{M}} (\ket{0} + e^{(2 \pi i u ) / 2^{1}} \ket{1}) \otimes (\ket{0} + e^{(2 \pi i u) / 2^{2} } \ket{1} ) \otimes  \dots \otimes  (\ket{0} + e^{( 2 \pi i u ) / 2^{l}} \ket{1}), \label{qft-eq-5}  
\eeqn
where $\tilde{u}$ and $v$ represent the quantum states in the Fourier basis and the computational basis, respectively, and $l$ denotes the number of qubits and $M$ is defined as $2^l$. The symbol $\otimes$ indicates the tensor product operation between qubits. For instance, $\ket{1} \otimes \ket{1} \otimes \ket{1} \otimes \ket{1} = \ket{1111} = \ket{15}$. 
In order to provide an illustration, we assume $l=4$ and $M=2^{4}$, and suppose the quantum state $\ket{\tilde{u}}$ is represented by $\ket{\tilde{15}}$ (i.e., $\ket{1111}$). Thus, in this case, the QFT can be expressed in~Eq.(\ref{qft-4-qubits}). The graphical representation of the QFT for $\ket{\tilde{15}}$ with 4 qubits, which maps between the computational basis and the Fourier basis, is depicted in Fig.~\ref{fig:qft-mapping-4-5-circuits}(a).
\begin{figure}[htb]
\beqn
\ket{\tilde{15}} &=&\frac{1}{\sqrt{16}} (\ket{0} + e^{(2 \pi i 15 ) / 2} \ket{1}) \otimes (\ket{0} + e^{(2 \pi i 15) / 4 } \ket{1} ) \otimes  (\ket{0} + e^{( 2 \pi i 15 ) / 8} \ket{1}) \label{qft-4-qubits} \nonumber \\
                      & &  \otimes (\ket{0} + e^{( 2 \pi i 15 ) / 16} \ket{1}). 
\eeqn
\end{figure}

\begin{figure*}[htb]
 \centering
 \captionsetup{justification=centering}
 \subfloat[QFT mapping.]{\label{fig:comp-to-qft-basis}\includegraphics[width=0.4\textwidth]{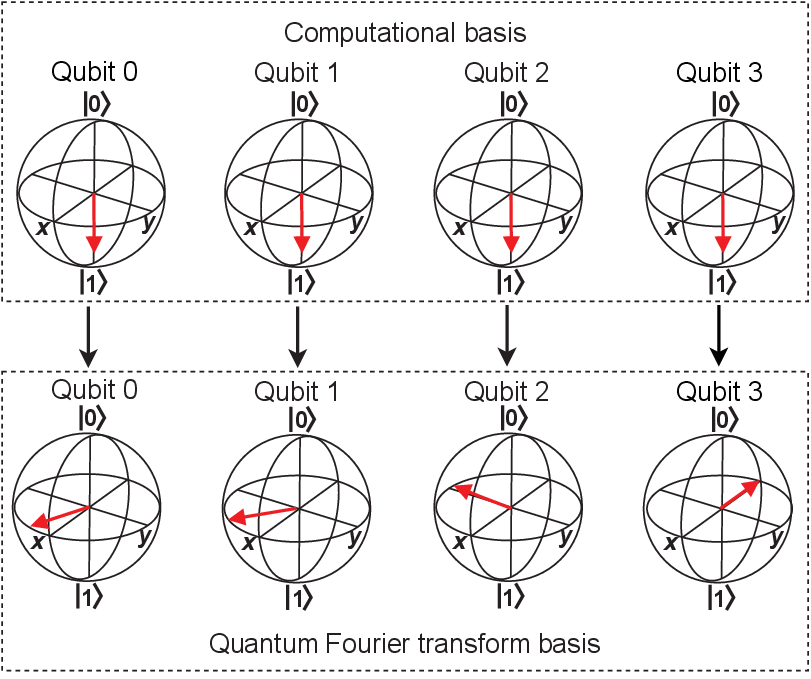}}
 \subfloat{\label{fig:qft-circuits}\includegraphics[width=0.6\textwidth]{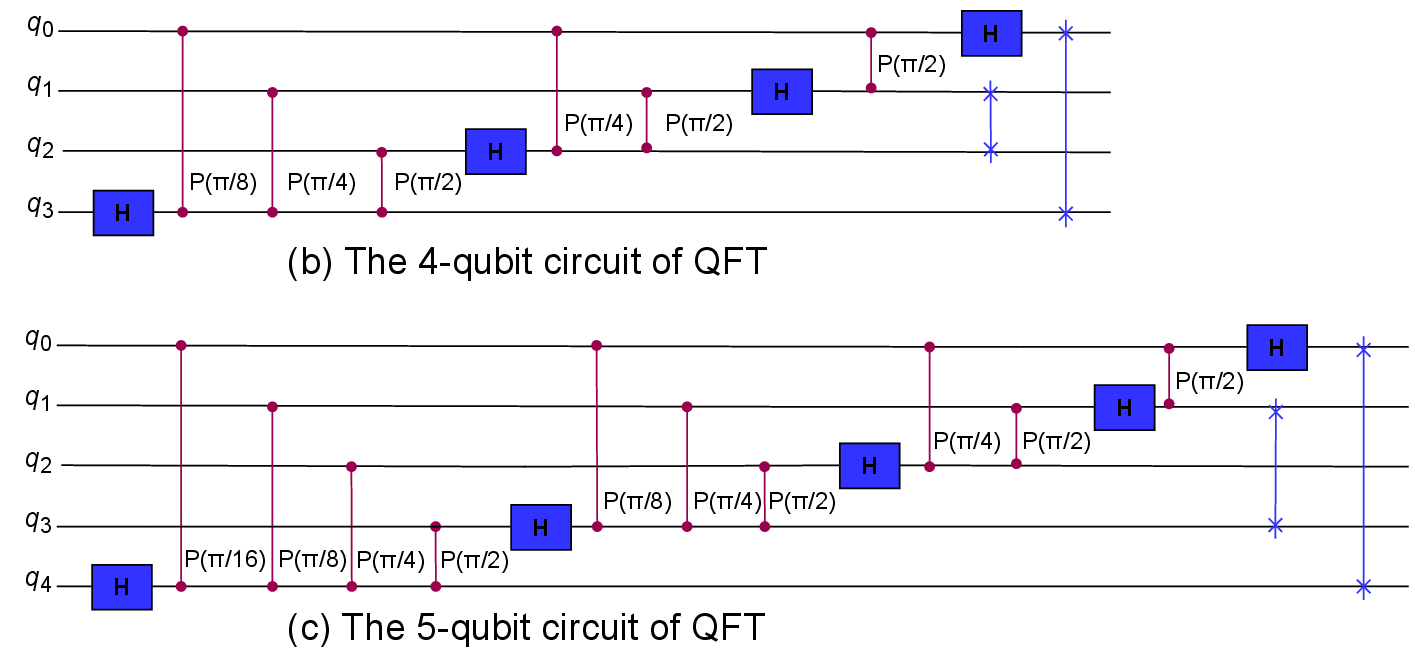}}
 \caption{(a) QFT mapping with 4 qubits, (b) 4-qubit circuit of QFT, and (c) 5-qubit quantum circuit of QFT.}
 \label{fig:qft-mapping-4-5-circuits}
\end{figure*} 

\subsubsection{QFT circuits by Qiskit} 
In Fig.~\ref{fig:qft-mapping-4-5-circuits}(b), we present two quantum circuits that implement QFT using 4 and 5 qubits. These circuits are created and visualized using Qiskit~\cite{quantum-fourier-transform-qiskit2022}. The circuits consist of qubits (i.e., $q_{0}$, $q_{1}$, $\dots$, $q_{l}$) and use several types of quantum gates: single-qubit Hadamard gate (a blue square with $\mathbf{H}$), two-qubit controlled rotation gate (a purple line), and SWAP gate (a blue line). For instance, in Fig.~\ref{fig:qft-mapping-4-5-circuits}(b), there is a quantum circuit that operates on 4 qubits ($q_{0}$ to $q_{3}$). The 4-qubit circuit performs a series of operations as follows. First, the $q_{3}$ qubit undergoes an $\mathbf{H}$-gate operation. Then, two $\mathbf{CROT}$-gates are used from $q_{3}$ to $q_{0}$, $q_{1}$ to $q_{3}$, and $q_{2}$ to $q_{3}$. Next, the $q_{2}$ qubit undergoes an $\mathbf{H}$-gate operation, followed by $\mathbf{CROT}$-gates from $q_{2}$ to $q_{0}$ and from $q_{1}$ to $q_{2}$. The $q_{1}$ qubit then undergoes an $\mathbf{H}$-gate operation, followed by a $\mathbf{CROT}$-gate from $q_{0}$ to $q_{1}$. Finally, the $q_{0}$ qubit undergoes an $\mathbf{H}$-gate operation, and SWAP gates apply for $q_{1}$ and $q_{2}$, and $q_{0}$ and $q_{3}$. 

Referring to Fig.~\ref{fig:qft-mapping-4-5-circuits}(b), it can be observed that as the number of qubits in a quantum circuit implementing QFT increases when the depth of the circuit and the number of quantum gates are used, resulting in increased computational time. This implies that the computational time required to execute QFT is directly proportional to the number of qubits and quantum gates involved, as described by Eq.~(\ref{qft-eq-5}) and illustrated in Fig.~\ref{fig:qft-mapping-4-5-circuits}(b).


\begin{table} [t]  \caption{\textcolor{black}{List of key notations for the SP model.}}
\label{table:notations-definitions}
\centering
\scalebox{0.9} {\color{black} \begin{tabular}{|l|l|}\hline
\bf{Notations} & \bf{Definitions} \\\hline 
    ${\mathcal{N}}$   & Set of all quantum nodes in the QCC \\ \hline
	${\mathcal{O}}_n$ & Set of outgoing links from node $n \in {\mathcal{N}}$ \\ \hline
    ${\mathcal{I}}_n$ & Set of incoming links to node $n \in {\mathcal{N}}$ \\ \hline
    ${\mathcal{R}}$   & Set of quantum requests \\ \hline
    $\mathcal{S}^{\mathrm{qc}}$ & Set of quantum circuits, $\mathcal{S}^{\mathrm{qc}} = \{1,\dots,c,\dots,C\}$ \\ \hline 
    $\mathcal{S}^{\mathrm{qp}}$ & Set of QCC providers, $\mathcal{S}^{\mathrm{qp}} = \{1,\dots,\rho,\dots,P\}$ \\ \hline
    $\mathcal{S}^{\mathrm{qm}}_{\rho}$ & Set of quantum computers in quantum provider $\rho$, $\mathcal{S}^{\mathrm{qm}}_{\rho} = \{1,\dots,m,\dots,M\}$ \\ \hline  
    $P^{\mathrm{wt}}_{c,\rho}$ & Penalty cost of over-waiting time of circuit $c$ in provider $\rho$ \\ \hline
    $F^{\mathrm{fts}}_{i,n}$  & Fidelity threshold \\ \hline 
    $R^{\mathrm{ep}}_{n,r}$ & Reservation cost of entangled pair of node $n$ \\ \hline
    $U^{\mathrm{ep}}_{n,r}$ & Utilized cost of entangled pair of node $n$ \\ \hline
    $O^{\mathrm{ep}}_{n,r}$ & On-demand cost of entangled pair of node $n$ \\ \hline
    $R^{\mathrm{cq}}_{c,\rho}$ & Reservation cost of qubit charged by provider $\rho$ \\ \hline
    $U^{\mathrm{cq}}_{c,\rho}$ & Utilized cost of qubit charged by provider $\rho$ \\ \hline
    $O^{\mathrm{cq}}_{c,\rho}$ & On-demand cost of qubit for circuit $c$ charged by provider $\rho$ \\ \hline
    $w_{i,j,r}$             & Binary variable that determines whether request $r \in \mathcal{R}$ will utilize the link \\
                            & between nodes $i$ and $j$ as part of its route or not  \\ \hline
    $y^{\mathrm{rep}}_{i,j,r}$ & Decision variable that represents the number of entangled pairs between \\
                               & nodes $i$ and $j$ in the reservation phase \\ \hline
    $y^{\mathrm{eep}}_{i,j,r,\psi}$ & Decision variable that represents the number of entangled pairs between \\
                                    & nodes $i$ and $j$ under scenario $\psi$ in the utilization phase \\ \hline
    $y^{\mathrm{oep}}_{i,j,r,\psi}$ & Decision variable that represents the number of entangled pairs between  \\ 
                                    & nodes $i$ and $j$ under scenario $\psi$ in the on-demand phase \\ \hline 
    $x^{\mathrm{rqt}}_{c,\rho,m,r}$  & Non-negative integer variable that represents the number of qubits for circuit $c$ of  \\ 
                                     & request $r$ executing on computer $m$ of provider $\rho$ in the reservation phase \\ \hline   
    $x^{\mathrm{uqt}}_{c,\rho,m,r,\psi}$ & Non-negative integer variable that represents the number of qubits used by circuit $c$ of \\
                                         & request $r$ executing on computer $m$ of provider $\rho$ under scenario $\psi$ in the utilization phase \\ \hline  
    $x^{\mathrm{oqt}}_{c,\rho,r,\psi}$   & Non-negative integer variable that represents the number of qubits used by circuit $c$ of  \\
                                         & request $r$ in provider $\rho$ under scenario $\psi$ in the on-demand phase \\ \hline
    $y^{\mathrm{owt}}_{c,\rho,m,r,\psi}$ & Positive real variable that represents the over-waiting time for circuit $c$ of request $r$  \\ 
                                         &  executing on computer $m$ in provider $\rho$ under scenario $\psi$  \\ \hline                              
\end{tabular}}  
\end{table}


\section{Entangled Pair and Qubit Resource Allocation Formulation}
\label{sec:optimization-formulation}

In this section, we first introduce the sets, constants, and decision variables of the SP model. Next, we present the entangled pair and qubit resource allocation based on the two-stage SP~\cite{Brige1997}, and provide the deterministic equivalent formulation for solving the SP model.

\subsection{Model Description}

The sets, constants, and decision variables of the SP model are described in Table \ref{table:notations-definitions}. We consider fidelity requirements, the number of qubits, and the expected execution time as uncertain parameters in the SP model. Let $\tilde{\psi}$ denote the composite random variable representing the requirements, which is expressed as follows: $\tilde{\psi} = \{ (\tilde{\delta}_{r,c},\tilde{\beta}_{r,c},\tilde{\alpha}_{r,c}) | \tilde{\delta}_{r,c} \in \mathcal{F}_{r,c}, \tilde{\beta}_{r,c} \in \mathcal{Q}_{r,c}, \tilde{\alpha}_{r,c} \in \mathcal{E}_{r,c} \}.$, where $\tilde{\delta}_{r,c},\tilde{\beta}_{r,c}$, and $\tilde{\alpha}_{r,c}$ are the random variables of fidelity value for circuit $c$ of request $r$, number of qubits required by circuit $c$ of request $r$, and expected execution time of circuit $c$ of request $r$, respectively. 

\subsection{Stochastic Programming Formulation}
\label{subsec:sp}
Our proposal is the two-stage SP model, which aims to provide entangled pair resources, ensure fidelity entanglement routing, and allocate qubit resources for running quantum circuits in the QCC, which is expressed as follows: 
\beqn
 \min_{w_{i,n,r}, y^{\mathrm{rep}}_{i,n,r}, x^{\mathrm{rqt}}_{c,\rho,m,r}}  & & \sum_{r \in {\mathcal{R}}} \Big(  \sum_{n \in {\mathcal{N}} } \sum_{i \in {\mathcal{I}}_n } ( ( E^{\mathrm{cc}}_{n} + S^{\mathrm{cs}}_{n} ) w_{i,n,r} y^{\mathrm{rep}}_{i,n,r} + R^{\mathrm{ep}}_{n,r} y^{\mathrm{rep}}_{i,n,r} ) \nonumber \\
	& &  + \sum_{c \in \mathcal{S}^{\mathrm{qc}}_r} \sum_{\rho \in \mathcal{S}^{\mathrm{qp}}} \sum_{m \in \mathcal{S}^{\mathrm{qm}}_{\rho}} x^{\mathrm{rqt}}_{c,\rho,m,r} R^{\mathrm{cq}}_{c,\rho}  \Big) + {\mathbb{E}}_{\Phi} \left[ {\mathscr{C}} ( w_{i,n,r}, y^{\mathrm{rep}}_{i,n,r}, x^{\mathrm{rqt}}_{c,\rho,m,r}, \tilde{\psi} ) \right]. 	\label{eq:sp_obj1} 
\eeqn
Theoretically, the SP model with the random variable $\tilde{\psi}$ in Eq. (\ref{eq:sp_obj1}) can be transformed into the deterministic equivalent formulation~\cite{Brige1997} which is expressed in Eqs. (\ref{eq:def_obj}) - (\ref{eq:def_const15}). Let $\mathcal{F}_{r,c}$ denote the set of the possible fidelity values for quantum circuit $c$ of request $r$, which is defined as $\mathcal{F}_{r,c} = \{ \delta_{r,c,\varrho} | \delta_{r,c,\varrho} \in [0.0, 1.0] \}$. Let $\mathcal{Q}_{r,c}$ denote the set of the potential number of qubits that a request $r$'s quantum circuit $c$ may require, which is defined as $\mathcal{Q}_{r,c} = \{ \beta_{r,c,1}, \beta_{r,c,2}, \dots, \beta_{r,c,{\iota}}\}$. Let $\mathcal{E}_{r,c}$ denote the set of the expected execution time for the quantum circuit $c$ of request $r$, which is defined as $\mathcal{E}_{r,c} = \{ \alpha_{r,c,1}, \alpha_{r,c,2}, \dots, \alpha_{r,c,\iota^{'}}\}$. $\iota$ and $\iota^{'}$ refer to the final indexes of the elements within the finite sets $\mathcal{Q}_{r,c}$ and $\mathcal{E}_{r,c}$, respectively. Let $\psi$ denote a scenario of request $r$. The scenario is a realization of the random variable $\tilde{\psi}$. Therefore, the potential value of the random variable can be selected from a set of scenarios. We use the term $\Phi$ to denote the collection of all scenarios, which we refer to as the \emph{scenario space}. The set of all scenarios of request $r$ is denoted as $\Psi_r$. The set of all scenarios is expressed as follows:
\beqn
    \Phi = {\displaystyle \prod_{r \in \mathcal{R}}}  \Psi_{r} = \Psi_{1} \times \Psi_{2} \times \dots \times \Psi_{|\mathcal{R}|}, \quad\quad\quad\quad\quad\quad\quad\quad\quad\quad\quad\quad\quad\quad\quad\; \\
    \mbox{where} \;\; \Psi_r = \mathcal{F}_{r,c} \times \mathcal{Q}_{r,c} \times \mathcal{E}_{r,c} = \{ (\delta_{r,c},\beta_{r,c},\alpha_{r,c}) |  \delta_{r,c} \in \mathcal{F}_{r}, \beta_{r,c} \in \mathcal{Q}_{r,c}, \alpha_{r,c} \in \mathcal{E}_{r,c} \}.
\eeqn
$\times$ and $|\mathcal{R}|$ are the Cartesian product and the cardinality of the set $\mathcal{R}$, respectively. Therefore, $\psi$ is the scenario space of request $r$ (i.e., $\psi \in \Psi_{r}$). The probability that requirements of fidelity, qubit and expected execution time of circuit $c$ of request $r$ are realized is denoted as $\mathbf{P}_{r}(\psi)$. The expectation  $\mathbb{E}_{\Phi}[\; \cdot \;]$ of the SP model in Eq. (\ref{eq:sp_obj1}) can be represented by the weighted sum of scenarios and their probabilities $\mathbf{P}_r(\psi)$. 

The objective function presented in Eq. (\ref{eq:def_obj}) is to minimize the overall cost of entangled pairs across all quantum nodes for entanglement establishment, the number of required qubits, and circuits' over-waiting time. The decision variables $y^{\mathrm{eep}}_{i,n,r,\psi}, y^{\mathrm{oep}}_{i,n,r,\psi}, x^{\mathrm{uqt}}_{c,\rho,m,r,\psi},  x^{\mathrm{oqt}}_{c,\rho,m,r,\psi}$, and $y^{\mathrm{owt}}_{c,\rho,m,r,\psi}$ rely on $\psi \in \Psi_r$ which means that demands' values are available when $\psi$ is observed. 

Equations~(\ref{eq:sp_const1}) and~(\ref{eq:sp_const2}) guarantee that source node $S^{\mathrm{q}}_r$ of request $r$  and destination node $D^{\mathrm{q}}_r$ of request $r$ have only one outgoing route and one incoming route, respectively. Equation~(\ref{eq:sp_const3}) ensures that the number of outgoing routes can be equal to the number of incoming routes for all quantum nodes of request $r$ except source and destination nodes. Equation~(\ref{eq:sp_const4}) ensures that there is only one outgoing route for the request $r$ of any node. Equation~(\ref{eq:sp_const5}) defines that the number of entangled pairs reserved at a link between nodes $i$ and $n$ in the reservation phase can be less than or equal to the maximum capacity of entangled pairs ($R^{\mathrm{etp}}_{i,j}$). 

Equation~(\ref{eq:def_const6}) ensures that the entangled pair utilization is not more than the entangled pair reservation at a link between nodes $i$ and $n$ of request $r$. Equations~(\ref{eq:def_const7}) and~(\ref{eq:def_const8}) guarantee that the sum of entangled pairs used in utilization and on-demand phases satisfies the fidelity requirement and the fidelity threshold, respectively. $\mathbf{F}^{\mathrm{dep}}(\cdot)$ in Eqs. (\ref{eq:def_const7}) and (\ref{eq:def_const8}) is the entanglement purification algorithm that is applied to enhance the entanglement fidelity based on the numbers of entangled pairs. Equation~(\ref{eq:def_const9}) defines the maximum capacity constraint of the on-demand phase ($C^{\mathrm{oep}}_{i,j}$) for utilizing the entangled pairs at a link between nodes $i$ and $j$. 

Equation~(\ref{eq:def_const12}) guarantees that  the number of qubits used in the utilization phase can be less than or equal to the qubits reserved in the reservation phase. Equation (\ref{eq:def_const13}) guarantees that the number of qubits in the utilization and on-demand phases can satisfy the qubit demands of quantum circuit $c$ for request $r$ ($\beta_{r,c,\psi}$). The requirement in Eq. (\ref{eq:def_const14}) is that quantum circuit $c$'s waiting time for request $r$ ($\alpha_{r,c,\psi}$) can be fulfilled. If quantum computer $m$ of provider $\rho$ takes more time to execute quantum circuit $c$ for request $r$ ($E^{\mathrm{exe}}_{c,\rho,m,r}$) than the waiting time of quantum circuit $c$, the additional waiting time of quantum circuit $c$ ($y^{\mathrm{owt}}_{c,\rho,m,r,\psi}$) will be charged. Equation (\ref{eq:sp_const15}) limits the number of qubits in the reservation phase for quantum circuits to be less than the maximum number of qubits of quantum machine $m$ of provider $\rho$ ($C^{\mathrm{qbt}}_{\rho,m}$). Equation~(\ref{eq:sp_const10}) defines that the decision variable is the binary integer. Equations~(\ref{eq:def_const11}) and~(\ref{eq:def_const15}) define that all decision variables are non-negative integers excluding $y^{\mathrm{owt}}_{c,\rho,m,r,\psi}$ that is the positive real number.

\beqn
	& & \min_{ w_{i,n,r}, y^{\mathrm{rep}}_{i,n,r}, y^{\mathrm{eep}}_{i,n,r,\psi}, y^{\mathrm{oep}}_{i,n,r,\psi}, x^{\mathrm{rqt}}_{c,\rho,m,r}, x^{\mathrm{uqt}}_{c,\rho,m,r,\psi}, x^{\mathrm{oqt}}_{c,\rho,m,r,\psi}, y^{\mathrm{owt}}_{c,\rho,m,r,\psi} } \label{eq:def_obj}  \nonumber \\
    & &	\sum_{r \in {\mathcal{R}}} \Big( \sum_{n \in {\mathcal{N}} } \sum_{i \in {\mathcal{I}}_n } ( ( E^{\mathrm{cc}}_{n} + S^{\mathrm{cs}}_{n} ) w_{i,n,r} y^{\mathrm{rep}}_{i,n,r} + R^{\mathrm{ep}}_{n,r} y^{\mathrm{rep}}_{i,n,r} ) + \sum_{c \in \mathcal{S}^{\mathrm{qc}}_r} \sum_{\rho \in \mathcal{S}^{\mathrm{qp}}} \sum_{m \in \mathcal{S}^{\mathrm{qm}}_{\rho}} x^{\mathrm{rqt}}_{c,\rho,m,r} R^{\mathrm{cq}}_{c,\rho}  \Big)  \nonumber \\
	& &  + \sum_{r \in {\mathcal{R}}}   \Big( \mathbf{P}_{r}(\psi) \big( \sum_{n \in {\mathcal{N}} } \sum_{i \in {\mathcal{I}}_n } ( U^{\mathrm{ep}}_{n, r} y^{\mathrm{eep}}_{i,n,r,\psi} + O^{\mathrm{ep}}_{n, r} y^{\mathrm{oep}}_{i,n,r,\psi}  ) \nonumber \\
    & & + \sum_{c \in \mathcal{S}^{\mathrm{qc}}_r} \sum_{\rho \in \mathcal{S}^{\mathrm{qp}}} \sum_{m \in \mathcal{S}^{\mathrm{qm}}_{\rho}} ( x^{\mathrm{uqt}}_{c,\rho,m,r,\psi} U^{\mathrm{cq}}_{c,\rho} + x^{\mathrm{oqt}}_{c,\rho,m,r,\psi} O^{\mathrm{cq}}_{c,\rho} + y^{\mathrm{owt}}_{c,\rho,m,r,\psi} P^{\mathrm{wt}}_{c,\rho})   \big) \Big),  \\ 
\mbox{s.t.} 
    & & \sum_{ k' \in {\mathcal{O}}_{S^{\mathrm{q}}_r} }	w_{S^{\mathrm{q}}_r,k',r}	-	\sum_{ h' \in {\mathcal{I}}_{S^{\mathrm{q}}_r} }	w_{h',S^{\mathrm{q}}_r,r} =	1,	 r	\in {\mathcal{R}},	\label{eq:sp_const1} \\
	& & \sum_{ h' \in {\mathcal{I}}_{D^{\mathrm{q}}_r} } w_{h', D^{\mathrm{q}}_r, r} - \sum_{ k' \in {\mathcal{O}}_{D^{\mathrm{q}}_r} } w_{D^{\mathrm{q}}_r, k', r } =	1,	 r \in {\mathcal{R}},	 \label{eq:sp_const2} \\
	& & \sum_{ k' \in {\mathcal{O}}_n } w_{n, k', r }	-	\sum_{h' \in {\mathcal{I}}_n } w_{h',n, r}	=	0,	 r	\in {\mathcal{R}}, n \in {\mathcal{N}} \setminus \{ S^{\mathrm{q}}_r, D^{\mathrm{q}}_r \}, \label{eq:sp_const3} \\
	& & \sum_{k' \in {\mathcal{O}}_n } w_{n, k', r } 	\leq	1,	 n \in {\mathcal{N}}, r \in {\mathcal{R}}, \label{eq:sp_const4} \\ 
    & & \sum_{r \in {\mathcal{R}}}  y^{\mathrm{rep}}_{i,n,r} w_{i,n, r} \leq R^{\mathrm{etp}}_{i,j}, i,j,n \in {\mathcal{N}}, \label{eq:sp_const5} \\ 
	& & y^{\mathrm{eep}}_{i,n,r,\psi} w_{i, n, r} \leq y^{\mathrm{rep}}_{i,n,r} w_{i,n,r}, i,j,n \in {\mathcal{N}}, r \in {\mathcal{R}}, \forall \psi \in \Psi_{r}, \label{eq:def_const6} \\
	& & \mathbf{F}^{\mathrm{dep}}\big( ( y^{\mathrm{eep}}_{i,n,r,\psi} w_{i,j, r}) + y^{\mathrm{oep}}_{i,n,r,\psi} \big) \geq \delta_{r,c,\psi}, i,n \in {\mathcal{N}}, r \in {\mathcal{R}}, \psi \in \Psi_{r}, \label{eq:def_const7} \\  
    & & \mathbf{F}^{\mathrm{dep}}\big( ( y^{\mathrm{eep}}_{i,n,r,\psi} w_{i,j, r}) + y^{\mathrm{oep}}_{i,n,r,\psi} \big) \geq F^{\mathrm{fts}}_{i,n}, i,n \in {\mathcal{N}}, r \in {\mathcal{R}}, \psi \in \Psi_{r}, \label{eq:def_const8}\\ 
    & & \sum_{r \in {\mathcal{R}}} \big( y^{\mathrm{oep}}_{i,j, r, \psi}  w_{i,j, r} \big) \leq C^{\mathrm{oep}}_{i,j}, i,j \in {\mathcal{N}}, \forall \psi \in \Psi_{r}, \label{eq:def_const9} \\
    & & x^{\mathrm{uqt}}_{c,\rho,m,r,\psi} \leq  x^{\mathrm{rqt}}_{c,\rho,m,r}, r \in {\mathcal{R}}, \forall c \in \mathcal{S}^{\mathrm{qc}}_r, \forall \rho \in \mathcal{S}^{\mathrm{qp}}, \forall m \in \mathcal{S}^{\mathrm{qm}}_{\rho}, \forall \psi \in \Psi_r, \label{eq:def_const12} \\    
	  & & x^{\mathrm{uqt}}_{c,\rho,m,r,\psi} + x^{\mathrm{oqt}}_{c,\rho,m,r,\psi} \geq  \beta_{r,c,\psi}, r \in {\mathcal{R}}, \forall c \in \mathcal{S}^{\mathrm{qc}}_r, \forall \rho \in \mathcal{S}^{\mathrm{qp}}, \forall m \in \mathcal{S}^{\mathrm{qm}}_{\rho}, \forall \psi \in \Psi_r,  \label{eq:def_const13} \\	
    & & E^{\mathrm{exe}}_{c,\rho,m,r} \leq  \alpha_{r,c,\psi} +  y^{\mathrm{owt}}_{c,\rho,m,r,\psi}, r \in {\mathcal{R}}, \forall c \in \mathcal{S}^{\mathrm{qc}}_r, \forall \rho \in \mathcal{S}^{\mathrm{qp}}, \forall m \in \mathcal{S}^{\mathrm{qm}}_{\rho}, \forall \psi \in \Psi_r,  \label{eq:def_const14} \\	
    & &  x^{\mathrm{rqt}}_{c,\rho,m,r} \leq C^{\mathrm{qbt}}_{\rho,m}, r \in {\mathcal{R}}, \forall c \in \mathcal{S}^{\mathrm{qc}}_r, \forall \rho \in \mathcal{S}^{\mathrm{qp}}, \forall m \in \mathcal{S}^{\mathrm{qm}}_{\rho}, \label{eq:sp_const15}  \\ 
    & & w_{i,n,r} \in \{ 0, 1 \}, i,n \in {\mathcal{N}}, r \in {\mathcal{R}},  \label{eq:sp_const10} \\
	& & y^{\mathrm{rep}}_{i, n, r }, y^{\mathrm{eep}}_{i, n, r, \psi }, y^{\mathrm{oep}}_{i, n, r, \psi } \in \mathbb{Z}^{\ast}, i,n \in {\mathcal{N}}, r \in {\mathcal{R}}, \forall \psi \in \Psi_{r},  \label{eq:def_const11} \\
	& & x^{\mathrm{rqt}}_{c,\rho,m,r}, x^{\mathrm{uqt}}_{c,\rho,m,r,\psi}, x^{\mathrm{oqt}}_{c,\rho,m,r,\psi} \in  \mathbb{Z}^{\ast}, y^{\mathrm{owt}}_{c,\rho,m,r,\psi} \in \mathbb{R}^{+}, \nonumber \\
    & & r \in {\mathcal{R}}, \forall c \in \mathcal{S}^{\mathrm{qc}}_r,  \forall \rho \in \mathcal{S}^{\mathrm{qp}}, \forall m \in \mathcal{S}^{\mathrm{qm}}_{\rho}, \forall \psi \in \Psi_r \label{eq:def_const15}.    
\eeqn

\section{Benders Decomposition}
\label{sec:benders-decomposition}

In this section, given the entanglement routing ($w^{\ast}_{i,n,r}$) for request $r$, we apply the Benders decomposition algorithm \cite{a-j-conejo2006} to solve the SP problem proposed in Section \ref{sec:optimization-formulation}. The objective of the algorithm is to divide the SP problem into multiple smaller SP problems that can be solved independently and concurrently. Note that this is solved on classical computers, while the algorithm is also applicable to quantum computers to solve, but we leave it as future work to optimize the solution algorithm for quantum computers. As a result, the complexity of the problem is reduced and the computational time to achieve the solution of the problem is shortened. The Benders decomposition algorithm can divide the SP problem presented in Eqs. (\ref{eq:def_obj}) - (\ref{eq:def_const15}) with complicating variables into {\em master problem} and {\em subproblem}. In the problem, we separately apply the Benders decomposition algorithm to the entangled pair resource allocation and the qubit resource allocation. 

\subsection{Entangled Pair Resource Allocation} 
In the entangled pair resource allocation, decision variables $y^{\mathrm{eep}}_{i,n,r,\psi}$ are the complicating variables. If variables $y^{\mathrm{eep}}_{i,n,r,\psi}$ are the fixed values and denoted as $y^{\mathrm{eepfix}}_{i,n,r,\psi}$, the entangled pair resource allocation can be decomposed into a master problem and two subproblems. Let $M^{\mathrm{ep}}$ denote the master problem which is presented as follows:
\beqn
	z^{\mathrm{eep}}_{\nu} & & = \min_{ y^{\mathrm{eep}}_{i,n,r,\psi,\nu} } \sum_{r \in {\mathcal{R}}} \mathbf{P}_{r}(\psi) \sum_{n \in {\mathcal{N}} } \sum_{i \in {\mathcal{I}}_n } ( U^{\mathrm{ep}}_{n, r} y^{\mathrm{eep}}_{i,n,r,\psi,\nu} ) + \alpha_{\nu}, \label{eq:obj_ep_master} \\ 
\mbox{s.t.} 
    & & \alpha_{\nu} \geq \alpha^{\mathrm{lwb}}_{\nu}, \label{eq:ep_master_const1} \\
	& & \mathbf{F}^{\mathrm{dep}}( y^{\mathrm{eep}}_{i,n,r,\psi,\nu} w^{\ast}_{i,j, r}) \leq \delta_{r,c,\psi}, i,n \in {\mathcal{N}}, r \in {\mathcal{R}}, \psi \in \Psi_{r}, \label{eq:ep_master_const2} \\  
    & & \mathbf{F}^{\mathrm{dep}}( y^{\mathrm{eep}}_{i,n,r,\psi,\nu} w^{\ast}_{i,j, r}) \leq F^{\mathrm{fts}}_{i,n}, i,n \in {\mathcal{N}}, r \in {\mathcal{R}}, \psi \in \Psi_{r}, \label{eq:ep_master_const3} \\ 
    & & \sum_{r \in {\mathcal{R}}} \sum_{\psi \in \Psi_{r}}  y^{\mathrm{eep}}_{i,n,r,\psi,\nu} w^{\ast}_{i,n, r} \leq R^{\mathrm{etp}}_{i,j}, i,j,n \in {\mathcal{N}}, \label{eq:ep_master_const4} \\ 
	& & y^{\mathrm{eep}}_{i, n, r, \psi,\nu} \in \mathbb{Z}^{\ast}, i,n \in {\mathcal{N}}, r \in {\mathcal{R}}, \forall \psi \in \Psi_{r}.  \label{eq:ep_master_const5}  
\eeqn
The objective function in Eq. (\ref{eq:obj_ep_master}) is derived from Eq. (\ref{eq:def_obj}). Let $\alpha_{\nu}$ denote the minimum costs of reservation and on-demand phases. $\alpha_{\nu}$ is updated in each iteration $\nu$. Let $\alpha^{\mathrm{lwb}}_{\nu}$ in Eq. (\ref{eq:ep_master_const1}) denote the lower bound of the minimum costs of reservation and on-demand phases. Let $\nu$ denote the iteration counter and initially set $\nu = 1$. Constraints in Eqs. (\ref{eq:ep_master_const2}) - (\ref{eq:ep_master_const5}) are the boundary of $y^{\mathrm{eep}}_{i,n,r,\psi,\nu}$. Let $S^{\mathrm{ep}}_{1}$ and $S^{\mathrm{ep}}_{2}(\psi)$ denote the subproblems  1 and 2, respectively. The subproblem $S^{\mathrm{ep}}_{1}$ is presented as follows:
\beqn
	z^{\mathrm{rep}}_{\nu} & & = \min_{ y^{\mathrm{rep}}_{i,n,r} } \sum_{r \in {\mathcal{R}}} \sum_{n \in {\mathcal{N}} } \sum_{i \in {\mathcal{I}}_n } ( E^{\mathrm{cc}}_{n} + S^{\mathrm{cs}}_{n} ) w_{i,n,r} y^{\mathrm{rep}}_{i,n,r} +  R^{\mathrm{ep}}_{n,r} y^{\mathrm{rep}}_{i,n,r},  \label{eq:ep_sub1_obj}  \\
\mbox{s.t.}
    & & (\ref{eq:sp_const5}), (\ref{eq:def_const6}), \; \mbox{and} \; (\ref{eq:def_const11}), \nonumber \\
    & & w_{i,n,r} = w^{\ast}_{i,n,r}, i,n \in {\mathcal{N}}, r \in {\mathcal{R}}, \label{eq:sub1_const1} \\
    & & y^{\mathrm{eep}}_{i,n,r,\psi} = y^{\mathrm{eepfix}}_{i,n,r,\psi}, i,n \in {\mathcal{N}}, r \in {\mathcal{R}}, \forall \psi \in \Psi_{r}. \label{eq:sub_entangled_pair_cost_fix} 
\eeqn     
The subproblem $S^{\mathrm{ep}}_{2}(\psi)$ is presented as follows:
\beqn
	z^{\mathrm{oep}}_{\nu}(\psi) &=& \min_{ y^{\mathrm{oep}}_{i,n,r,\psi} } \sum_{r \in {\mathcal{R}}}  \mathbf{P}_{r}(\psi) \sum_{n \in {\mathcal{N}} } \sum_{i \in {\mathcal{I}}_n } U^{\mathrm{ep}}_{n, r} y^{\mathrm{eep}}_{i,n,r,\psi} \label{eq:ep_sub2_obj} + O^{\mathrm{ep}}_{n, r} y^{\mathrm{oep}}_{i,n,r,\psi},     \\
\mbox{s.t.} 
    & & (\ref{eq:def_const7}), (\ref{eq:def_const8}), (\ref{eq:def_const9}), (\ref{eq:sub1_const1}), \; \mbox{and} \; (\ref{eq:sub_entangled_pair_cost_fix}). \nonumber  
\eeqn

\subsection{Qubit Resource Allocation} 
In qubit resource allocation, decision variables $x^{\mathrm{uqt}}_{c,\rho,m,r,\psi}$ are the complicating variables. If variables $x^{\mathrm{uqt}}_{c,\rho,m,r,\psi}$ have fixed values and denoted as $x^{\mathrm{uqtfix}}_{c,\rho,m,r,\psi}$, the qubit resource allocation can be decomposed into two subproblems, namely, $S^{\mathrm{qt}}_{1}$ and $S^{\mathrm{qt}}_{2}$. Let $M^{\mathrm{qt}}$ denote the master problem which is presented as follows:
\beqn
    z^{\mathrm{uqt}}_{\acute{\nu}} &=& \min_{x^{\mathrm{uqt}}_{c,\rho,m,r,\psi,\acute{\nu}} } \sum_{r \in {\mathcal{R}}} \mathbf{P}_{r}(\psi) \sum_{c \in \mathcal{S}^{\mathrm{qc}}_r} \sum_{\rho \in \mathcal{S}^{\mathrm{qp}}} \sum_{m \in \mathcal{S}^{\mathrm{qm}}_{\rho}} ( x^{\mathrm{uqt}}_{c,\rho,m,r,\psi,\acute{\nu}} U^{\mathrm{cq}}_{c,\rho} ) + \theta_{\acute{\nu}},  \label{eq:qt_master_master} \\ 
\mbox{s.t.} 
    & & \theta_{\acute{\nu}} \geq \theta^{\mathrm{lwb}}_{\acute{\nu}}, \label{eq:qt_master_const1} \\ 
	& & x^{\mathrm{uqt}}_{c,\rho,m,r,\psi,\acute{\nu}} \leq  \beta_{r,c,\psi}, r \in {\mathcal{R}}, \forall c \in \mathcal{S}^{\mathrm{qc}}_r, \forall \rho \in \mathcal{S}^{\mathrm{qp}}, \forall m \in \mathcal{S}^{\mathrm{qm}}_{\rho}, \forall \psi \in \Psi_r, \label{eq:qt_master_const2} \\	
    & &  x^{\mathrm{uqt}}_{c,\rho,m,r,\psi,\acute{\nu}} \leq C^{\mathrm{qbt}}_{\rho,m}, r \in {\mathcal{R}}, \forall c \in \mathcal{S}^{\mathrm{qc}}_r, \forall \rho \in \mathcal{S}^{\mathrm{qp}}, \forall m \in \mathcal{S}^{\mathrm{qm}}_{\rho}, \forall \psi \in \Psi_r, \label{eq:qt_master_const3} \\
	& &  x^{\mathrm{uqt}}_{c,\rho,m,r,\psi,\acute{\nu}} \in  \mathbb{Z}^{\ast}, r \in {\mathcal{R}}, \forall c \in \mathcal{S}^{\mathrm{qc}}_r, \forall \rho \in \mathcal{S}^{\mathrm{qp}}, \forall m \in \mathcal{S}^{\mathrm{qm}}_{\rho}, \forall \psi \in \Psi_r. \label{eq:qt_master_const4}
\eeqn
The objective function in Eq. (\ref{eq:qt_master_master}) is derived from that in Eq. (\ref{eq:def_obj}). Let $\theta_{\acute{\nu}}$ denote the minimum costs of reservation and on-demand phases. $\theta_{\acute{\nu}}$ is updated in each iteration $\acute{\nu}$. Let $\theta^{\mathrm{lwb}}_{\acute{\nu}}$ in Eq. (\ref{eq:qt_master_const1}) denote the lower bound of the minimum costs of reservation and on-demand phases. Let $\acute{\nu}$ denote the iteration counter and initially set $\acute{\nu} = 1$. Constraints in Eqs. (\ref{eq:qt_master_const2}) - (\ref{eq:qt_master_const4}) are the boundary of $x^{\mathrm{uqt}}_{c,\rho,m,r,\psi,\acute{\nu}}$. The subproblem $S^{\mathrm{qt}}_{1}$ is presented as follows:
\beqn
	z^{\mathrm{rqt}}_{\acute{\nu}} &=& \min_{x^{\mathrm{rqt}}_{c,\rho,m,r}} \sum_{r \in {\mathcal{R}}}  \sum_{c \in \mathcal{S}^{\mathrm{qc}}_r} \sum_{\rho \in \mathcal{S}^{\mathrm{qp}}} \sum_{m \in \mathcal{S}^{\mathrm{qm}}_{\rho}} x^{\mathrm{rqt}}_{c,\rho,m,r} R^{\mathrm{cq}}_{c,\rho}, \label{eq:qt_sub1_obj}  \\ 
\mbox{s.t.} 
	& & (\ref{eq:sp_const15}) \; \mbox{and} \; (\ref{eq:def_const12}), \\
    & & x^{\mathrm{uqt}}_{c,\rho,m,r,\psi} = x^{\mathrm{uqtfix}}_{c,\rho,m,r,\psi}, r \in {\mathcal{R}}, \forall c \in \mathcal{S}^{\mathrm{qc}}_r, \forall \rho \in \mathcal{S}^{\mathrm{qp}}, \forall m \in \mathcal{S}^{\mathrm{qm}}_{\rho}, \forall \psi \in \Psi_r. \label{eq:sub_qubit_const_fix} 
\eeqn      
The subproblem $S^{\mathrm{qt}}_{2}(\psi)$ is presented as follows:
\beqn
	z^{\mathrm{oqt}}_{\acute{\nu}}(\psi) &=& \min_{x^{\mathrm{oqt}}_{c,\rho,m,r,\psi}, y^{\mathrm{owt}}_{c,\rho,m,r,\psi}} \sum_{r \in {\mathcal{R}}} \mathbf{P}_{r}(\psi) \sum_{c \in \mathcal{S}^{\mathrm{qc}}_r} \sum_{\rho \in \mathcal{S}^{\mathrm{qp}}} \sum_{m \in \mathcal{S}^{\mathrm{qm}}_{\rho}}  ( x^{\mathrm{uqt}}_{c,\rho,m,r,\psi} U^{\mathrm{cq}}_{c,\rho} + x^{\mathrm{oqt}}_{c,\rho,m,r,\psi} O^{\mathrm{cq}}_{c,\rho} \nonumber \\ 
    & & + y^{\mathrm{owt}}_{c,\rho,m,r,\psi} P^{\mathrm{wt}}_{c,\rho} ),   \label{eq:qt_sub2_obj} \\ 
\mbox{s.t.} 
    & & (\ref{eq:def_const13}), (\ref{eq:def_const14}), \; \mbox{and} \; (\ref{eq:sub_qubit_const_fix}). \nonumber
\eeqn

\subsection{Benders Decomposition Algorithm} 

The algorithm outlines a four-step approach for addressing the challenges of entangled pair resource allocation and qubit resource allocation problems.

{\em Step 1: Initialization of master problems.} This step initializes the master problems and performs only one time while {\em steps 2}, {\em 3}, and {\em 4} repeat in the algorithm. In this step, the master problems of the entangle pair resource allocation and qubit resource allocation, which are respectively expressed in Eqs. (\ref{eq:obj_ep_master}) - (\ref{eq:ep_master_const5}) and Eqs. (\ref{eq:qt_master_master}) - (\ref{eq:qt_master_const4}) are the alternative form of the deterministic equivalent formulation represented in Eqs.~(\ref{eq:def_obj}) - (\ref{eq:def_const15}).  

{\em Step 2: Subproblem solutions.} The subproblems $S^{\mathrm{ep}}_{1}$, $S^{\mathrm{ep}}_{2}(\psi)$, $S^{\mathrm{qt}}_{1}$, and $S^{\mathrm{qt}}_{2}(\psi)$ are formulated and solved. We assign solution $y^{\mathrm{eep}}_{i,n,r,\psi,\nu}$ obtained from the master problem in Eqs.~(\ref{eq:obj_ep_master}) - (\ref{eq:ep_master_const5}) to $y^{\mathrm{eepfix}}_{i,n,r,\psi}$ (i.e., $y^{\mathrm{eepfix}}_{i,n,r,\psi} = y^{\mathrm{eep}}_{i,n,r,\psi,\nu}$). Then, we assign solution $x^{\mathrm{uqt}}_{c,\rho,m,r,\psi,\acute{\nu}}$ obtained from the master problem in Eqs.~ (\ref{eq:qt_master_master}) - (\ref{eq:qt_master_const4}) to $x^{\mathrm{uqtfix}}_{c,\rho,m,r,\psi}$ (i.e., $x^{\mathrm{uqtfix}}_{c,\rho,m,r,\psi} = x^{\mathrm{uqt}}_{c,\rho,m,r,\psi,\acute{\nu}}$). Therefore, given the fixed solutions $y^{\mathrm{eepfix}}_{i,n,r,\psi}$ and $x^{\mathrm{uqtfix}}_{c,\rho,m,r,\psi}$, the subproblems $S^{\mathrm{ep}}_{1}$, $S^{\mathrm{ep}}_{2}(\psi)$, $S^{\mathrm{qt}}_{1}$, and $S^{\mathrm{qt}}_{2}(\psi)$ can be solved concurrently. 

The objective of subproblem $S^{\mathrm{ep}}_{1}$ in Eqs.~(\ref{eq:ep_sub1_obj})- (\ref{eq:sub_entangled_pair_cost_fix}) is to minimize the reservation cost regarding the entangled pairs. Let $\lambda^{\mathrm{rep}}_{i,n,r,\nu}$ denote the solution of the dual problem of $S^{\mathrm{ep}}_{1}$ in iteration $\nu$ associated with Eq.~(\ref{eq:sub_entangled_pair_cost_fix}). The objective of subproblem $S^{\mathrm{ep}}_{2}(\psi)$ in Eq.~(\ref{eq:ep_sub2_obj}) is to minimize the on-demand cost regarding the entangled pairs. Let $\lambda^{\mathrm{oep}}_{i,n,r,\nu}(\psi)$ denote the solution of the dual problem of $S^{\mathrm{ep}}_{2}(\psi)$ when the fidelity value (i.e., $\delta_{r,c,\varrho}$) is observed and set to $\psi$. The subproblem $S^{\mathrm{ep}}_{2}(\psi)$ associates with the number of scenarios $|\Psi^{\mathrm{ep}}_{r}|, \psi \in \Psi^{\mathrm{ep}}_{r} $ where $|\Psi^{\mathrm{ep}}_{r}|$ is the Cardinality of set $\Psi^{\mathrm{ep}}_{r}$, and therefore $|\Psi^{\mathrm{ep}}_{r}|$ is generated. Let $\lambda^{\mathrm{oep}}_{i,n,r,\nu}(\psi)$ denote the solution of the dual problem of $S^{\mathrm{ep}}_{2}(\psi)$ in iteration $\nu$ associated with Eq.~(\ref{eq:sub_entangled_pair_cost_fix}).

The objective of subproblem $S^{\mathrm{qt}}_{1}$ is to minimize the reservation cost of the qubit allocation. Let $\lambda^{\mathrm{qrt}}_{c,\rho,m,r,\acute{\nu}}$ denote the solution of the dual problem of $S^{\mathrm{qt}}_{1}$ in iteration $\acute{\nu}$ associated with Eq.~(\ref{eq:sub_qubit_const_fix}). The objective of subproblem $S^{\mathrm{qt}}_{2}(\psi)$ is to minimize the on-demand cost of the qubit allocation. Let $\lambda^{\mathrm{oqt}}_{c,\rho,m,r,\acute{\nu}}(\psi)$ denote the solution of the dual problem of $S^{\mathrm{qt}}_{2}(\psi)$ when the number of required qubits (i.e., $\beta_{r,c,{\iota}}$), and the waiting time for quantum circuits (i.e.,  $\alpha_{r,c,\iota^{'}}$) are observed and set to $\psi$. The subproblem $S^{\mathrm{qt}}_{2}(\psi)$ associates with the number of scenarios $|\Psi^{\mathrm{qt}}_{r}|, \psi \in \Psi^{\mathrm{qt}}_{r} $ where $|\Psi^{\mathrm{qt}}_{r}|$ is the Cardinality of set $\Psi^{\mathrm{qt}}_{r}$, and therefore $|\Psi^{\mathrm{qt}}_{r}|$ is generated. Let $\lambda^{\mathrm{oqt}}_{c,\rho,m,r,\acute{\nu}}(\psi)$ denote the solution of the dual problem of $S^{\mathrm{qt}}_{2}(\psi)$ in iteration $\acute{\nu}$ associated with Eq.~(\ref{eq:sub_qubit_const_fix}). The solutions of $\lambda^{\mathrm{rep}}_{i,n,r,\nu}$, $\lambda^{\mathrm{oep}}_{i,n,r,\nu}(\psi)$, $\lambda^{\mathrm{qrt}}_{c,\rho,m,r,\acute{\nu}}$, and $\lambda^{\mathrm{oqt}}_{c,\rho,m,r,\acute{\nu}}(\psi)$ will be applied in {\em step 4}.

{\em Step 3: Convergence checking.} The convergence of lower and upper bounds of solutions from master problems and subproblems are checked. Let $z^{\mathrm{lwb}}_{\nu}$ denote the lower bound in iteration $\nu$ obtained from the master problem in Eq. (\ref{eq:obj_ep_master}), which is $z^{\mathrm{lwb}}_{\nu}$ = $z^{\ast\mathrm{eep}}_{\nu}$. Let $z^{\mathrm{upb}}_{\nu}$ denote the upper bound in iteration $\nu$ that is obtained from $z^{\mathrm{upb}}_{\nu} = z^{\ast\mathrm{eep}}_{\nu} - \alpha_{\nu} + z^{\mathrm{rep}}_{\nu} + z^{\mathrm{oep}}_{\nu}$. Let $z^{\mathrm{lwb}}_{\acute{\nu}}$ denote the lower bound in iteration $\acute{\nu}$ obtained from the master in Eq. (\ref{eq:qt_master_master}), which is  $z^{\mathrm{lwb}}_{\acute{\nu}}$ = $z^{\ast \mathrm{uqt}}_{\acute{\nu}}$. Let $z^{\mathrm{upb}}_{\acute{\nu}}$ denote the upper bound in iteration $\acute{\nu}$ that is obtained from $z^{\mathrm{upb}}_{\acute{\nu}} = z^{\ast \mathrm{uqt}}_{\acute{\nu}} - \theta_{\acute{\nu}} + z^{\mathrm{rqt}}_{\acute{\nu}} + z^{\mathrm{oqt}}_{\acute{\nu}}$. Let $\epsilon$ and $\varepsilon$ denote the small tolerance values to verify the convergence of lower and upper bounds for entangled pair resource allocation and qubit resource allocation, respectively. If  $z^{\mathrm{upb}}_{\nu} - z^{\mathrm{lwb}}_{\nu} < \epsilon$ and  $z^{\mathrm{upb}}_{\acute{\nu}} - z^{\mathrm{lwb}}_{\acute{\nu}} < \varepsilon$, the Benders decomposition algorithm stops and the optimal solutions achieve. Otherwise, the algorithm performs to  {\em step 4}. 

 {\em Step 4: Master problem solutions.} 
\beqn
	\alpha_{\nu} &\geq& \sum_{r \in {\mathcal{R}}} \sum_{\psi \in \Psi} \sum_{n \in {\mathcal{N}} } \sum_{i \in {\mathcal{I}}_n } \Big( (\lambda^{\mathrm{rep}}_{i,n,r,\bar{\nu}} + \lambda^{\mathrm{oep}}_{i,n,r,\bar{\nu}}(\psi)) ( y^{\mathrm{eep}}_{i,n,r,\psi,\nu} - y^{\mathrm{eep}}_{i,n,r,\psi,\bar{\nu}} ) \Big) \nonumber \\
    & & + z^{\mathrm{\ast rep}}_{\bar{\nu}} +  \sum_{\psi \in \Psi} z^{\mathrm{\ast oep}}_{\bar{\nu}}(\psi), \bar{\nu} \in \{1,\dots,\nu -1\}, \label{eq:entangled-pair-bender-cuts} \\
    \theta_{\acute{\nu}} &\geq& \sum_{r \in {\mathcal{R}}} \sum_{\psi \in \Psi} \sum_{c \in \mathcal{S}^{\mathrm{qc}}_r} \sum_{\rho \in \mathcal{S}^{\mathrm{qp}}} \sum_{m \in \mathcal{S}^{\mathrm{qm}}_{\rho}} \Big( (\lambda^{\mathrm{qrt}}_{c,\rho,m,r,\ddot{\nu}} + \lambda^{\mathrm{oqt}}_{c,\rho,m,r,\ddot{\nu}}(\psi)) (x^{\mathrm{uqt}}_{c,\rho,m,r,\psi,\acute{\nu}} - x^{\mathrm{uqt}}_{c,\rho,m,r,\psi,\ddot{\nu}}) \Big) \nonumber \\ 
    & & + z^{\mathrm{\ast rqt}}_{\ddot{\nu}} + \sum_{\psi \in \Psi} z^{\mathrm{\ast oqt}}_{\ddot{\nu}}(\psi), \ddot{\nu} \in \{1,\dots,\acute{\nu} -1\}. \label{eq:qubit-bender-cuts}
\eeqn

The iteration counters $\nu$ and $\acute{\nu}$ are respectively incremented by $\nu = \nu + 1$ and $\acute{\nu} = \acute{\nu} + 1$. Then, the master problems of the entangled pair resource allocation in Eqs. (\ref{eq:obj_ep_master}) - (\ref{eq:ep_master_const5}) and qubit resource allocation in Eqs. (\ref{eq:qt_master_master}) - (\ref{eq:qt_master_const4}) can be relaxed by additional constraints (i.e., \emph{ Bender cuts}~\cite{a-j-conejo2006}). The solutions of the master problems update the costs $\alpha_{\nu}$ and $\theta_{\acute{\nu}}$ and the utilizing costs according to the solutions of $y^{\mathrm{eep}}_{i,n,r,\psi,\nu}$ and $x^{\mathrm{uqt}}_{c,\rho,m,r,\psi,\acute{\nu}}$. Benders cuts as shown in Eqs. (\ref{eq:entangled-pair-bender-cuts}) and (\ref{eq:qubit-bender-cuts}) are created from the optimal costs obtained from master problems and subproblems in the previous iterations. Once the master problems are solved, {\em step 2} is executed, and the iterative process is repeated.

\section{Performance Evaluation}
\label{sec:performance-evaluation} 

\subsection{Parameter Setting}

We evaluate the QCC system as illustrated in Fig.~\ref{fig:system-model}. The system consists of three QCC providers ($\mathcal{S}^{\mathrm{qp}} = \{1, 2, 3\}$), each of which has two quantum computers ($\mathcal{S}^{\mathrm{qm}}_\rho = \{1, 2\}$)~\cite{r-kaewpuang-stochastic-qubit2023}. For each provider $\rho$, we establish a maximum limit of 30 qubits~\cite{r-kaewpuang-stochastic-qubit2023} per quantum computer ($C^{\mathrm{qbt}}_{\rho,m}$). Initially, we assign the penalty cost of \$10~\cite{r-kaewpuang-stochastic-qubit2023} for the waiting time of circuit program $c$ when processed by provider $\rho$ ($P^{\mathrm{wt}}_{c,\rho}$). For qubit cost values charged by provider $\rho$ for circuit $c$, we set the reservation ($R^{\mathrm{cq}}_{c,\rho}$), utilization ($U^{\mathrm{cq}}_{c,\rho}$), and on-demand costs ($O^{\mathrm{cq}}_{c,\rho}$) to be \$1.68~\cite{ibm-quantum-computing}, \$0.1~\cite{r-kaewpuang-stochastic-qubit2023}, \$7~\cite{r-kaewpuang-stochastic-qubit2023}, respectively. We consider quantum circuits of quantum discrete Fourier transform (QDFT)~\cite{quantum-fourier-transform-qiskit2022} with different numbers of qubits, which we exemplify and perform. We consider that circuit program $c$ requires a random number of qubits between 10 and 22~\cite{r-kaewpuang-stochastic-qubit2023}, denoted by $\beta_{r,c,\psi} \in \{10,\dots,22\}$, and has a random waiting time between 0.001 and 0.009 seconds~\cite{r-kaewpuang-stochastic-qubit2023}, denoted as $\alpha_{r,c,\psi} \in \{0.001,\dots,0.009\}$. Both of these random variables are assumed to follow a uniform distribution. We conduct experiments on the NSFNET network connected via optical fibers~\cite{y-cao-hybrid-trusted-untrusted2021}. The initial values of fidelity between nodes $i$ and $j$ in the network are presented in Fig.~\ref{fig:routing-entangled-pairs-and-optimal-solution}(a). The initial fidelity threshold ($F^{\mathrm{fts}}_{i,n}$) is 0.8~\cite{j-li-fidelity-guaranteed-entanglement2022}. The largest quantity of entangled pairs between nodes $i$ and $j$ in the reservation ($R^{\mathrm{etp}}_{i,j}$) and the on-demand ($C^{\mathrm{oep}}_{i,j}$) phases are 9 and 60, respectively. The random quantity of fidelity requirements is set between 0.55 to 1.0 (i.e., $\delta_{r,c,\psi} \in \{0.55,\dots,1.0\}$) with uniform distribution. We also set the cost of an entangled pair for reservation ($R^{\mathrm{ep}}_{n,r}$), utilization ($U^{\mathrm{ep}}_{n, r}$), and on-demand ($O^{\mathrm{ep}}_{n, r}$) phases to \$10, \$1, and \$200, respectively~\cite{r-kaewpuang-entangled-pair-2023}. Additionally, energy consumption cost of node $n$ ($E^{\mathrm{cc}}_{n}$) is \$5 while energy consumption cost to establish repeater node $n$ ($S^{\mathrm{cs}}_{n}$) is \$151. For Benders decomposition, we set the small tolerance values to verify the convergence of lower and upper bounds for entangled pair resource allocation ($\epsilon$) and qubit resource allocation ($\varepsilon$) to be 0.05. We use the GAMS/CPLEX solver~\cite{Gams} to implement and solve the stochastic programming formulation.

\subsection{Numerical Results}

We divide the experiment into two parts: entangled pair resource allocation and qubit resource allocation. In the entangled pair resource allocation part, we conduct networking experiments based on the allocation of entangled pairs. This includes considerations such as entanglement routing, fidelity requirements, and entanglement purification. In the qubit resource allocation part, we perform computing experiments based on the allocation of computing qubits, quantum circuits, quantum computers, and QCC providers.

\subsubsection{Entangled pair resource allocation}

\begin{figure*}[htb]
 \centering
 \captionsetup{justification=centering}
 \subfloat[Route solutions for three requests.]{\label{fig:routing-three-requests}\includegraphics[width=0.33\textwidth]{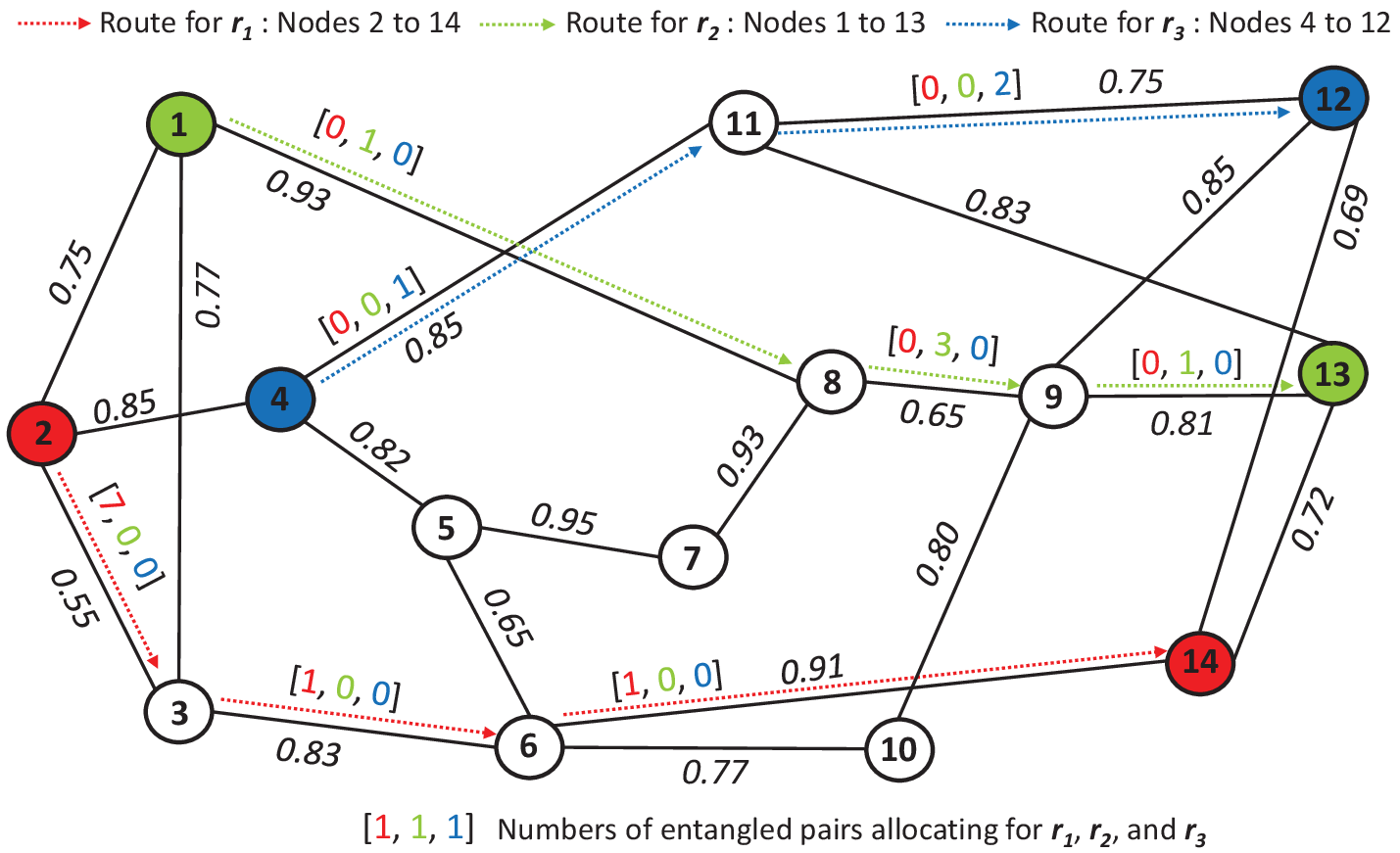}}
 \subfloat[Utilization of the entangled pairs in SP model.]{\label{fig:reserved-utilized-on-demand-entangled-pairs}\includegraphics[width=0.33\textwidth]{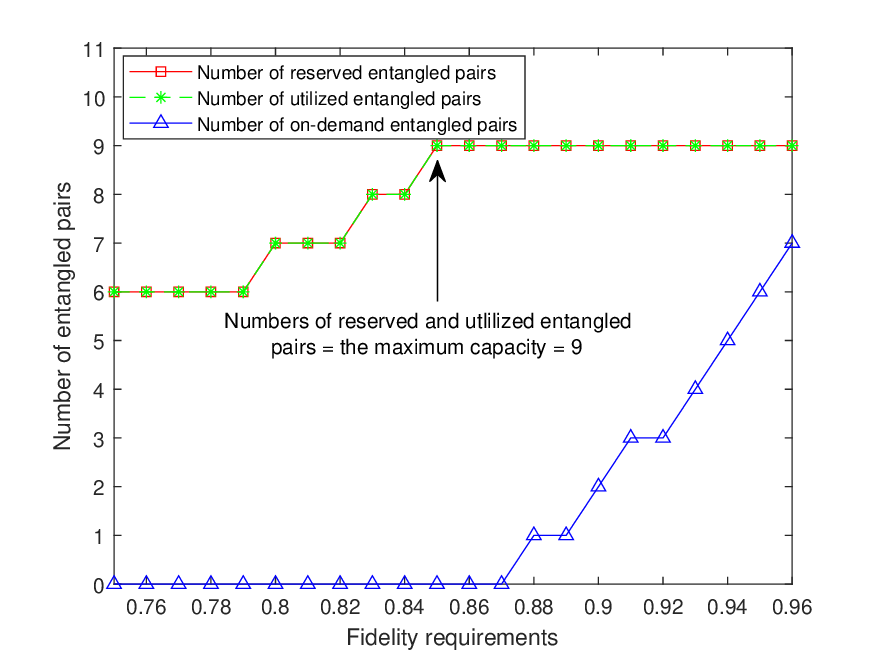}}
 \subfloat[The total and two-stage costs.]{\label{fig:optimal-solution-different-entangled-pairs}\includegraphics[width=0.33\textwidth]{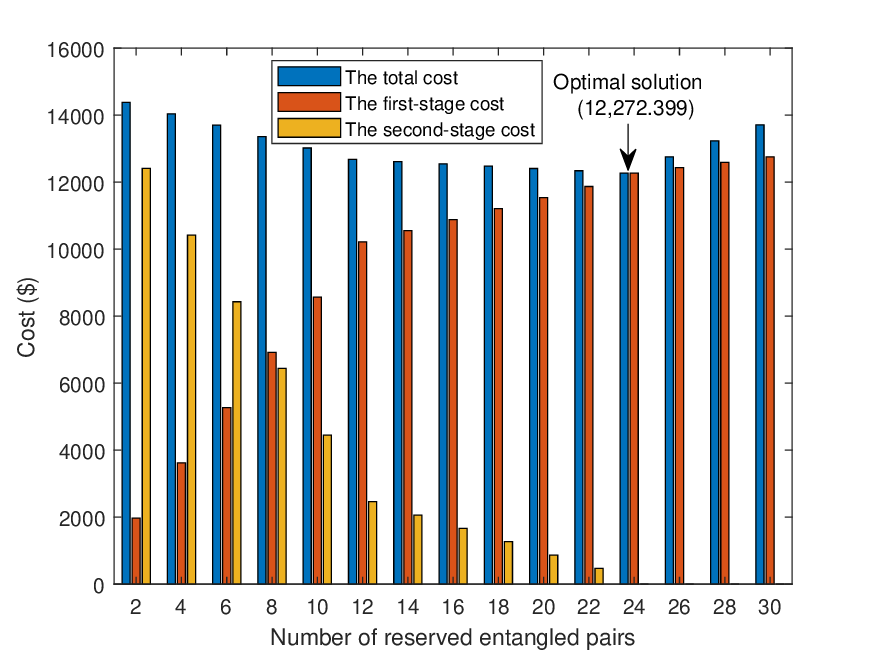}}
 \caption{(a) Requests $r_1$, $r_2$, and $r_3$ in NSFNET topology, (b) Comparison of entangled pair utilization of 3 phases, and (c) The optimal solution in SP model.}
 \label{fig:routing-entangled-pairs-and-optimal-solution}
 \vspace{-0.5cm}
\end{figure*} 


Figure~\ref{fig:routing-entangled-pairs-and-optimal-solution}(a) shows the solutions generated by the proposed model that meet the fidelity requirements of three distinct requests, namely $r_1$, $r_2$, and $r_3$. For each request, the SP model determines the optimal route and the number of entangled pairs needed to comply with the fidelity requirement in the network. For example, the route for request $r_1$ is nodes $2 \rightarrow 3 \rightarrow 6 \rightarrow 14$ and the SP model allocates 9 entangled pairs to satisfy its fidelity requirement. Additionally, we observe that the number of entangled pairs between nodes 2 and 3 (i.e., $2 \rightarrow 3$) is utilized at 7 since the fidelity value between these nodes (i.e., $0.55$) is less than the fidelity requirement (i.e., $0.80$). Therefore, the entangled pairs are more utilized for entanglement purification to enhance the fidelity value and satisfy the fidelity requirement. 

Figure~\ref{fig:routing-entangled-pairs-and-optimal-solution}(b) illustrates the number of entangled pairs in three different phases (reservation, utilization, and on-demand) across various fidelity requirements. As depicted in the figure, the quantity of reserved and utilized entangled pairs rises gradually in the reservation and utilization phases until the fidelity requirement reaches 0.87. Once this value is reached, the reserved and utilized entangled pairs reach their peak capacity of 9 pairs and are unable to accommodate higher fidelity requirements. Hence, to fulfill more demanding fidelity requirements, the entangled pairs in the on-demand phase are used. In this phase, the utilization of entangled pairs begins at a fidelity requirement of 0.88, as the reservation phase has limited capacity for entangled pairs.

Figure~\ref{fig:routing-entangled-pairs-and-optimal-solution}(c) demonstrates the effectiveness of the SP model in achieving the optimal solution. We vary the quantity of reserved entangled pairs and show the optimal outcome achieved through the model, along with the influence of reserved entangled pairs on the solution. As shown in Fig.~\ref{fig:routing-entangled-pairs-and-optimal-solution}(c), the first-stage cost rises significantly as the number of reserved entangled pairs increases. However, the second-stage cost decreases significantly after the fidelity requirements are met. This is because the number of entangled pairs in the reservation phase is constrained to be maximum due to the lower cost, while the number of entangled pairs in the on-demand phase is constrained to be minimum. Therefore, the optimal solution is reached at 24 reserved entangled pairs, with a cost of \$12,272.399 and the second-stage cost of \$0, since the reserved entangled pairs meet the fidelity requirements, and there is no need to utilize on-demand entangled pairs in the second stage. After 24 reserved entangled pairs, both the total cost and the first-stage cost slightly rise because of the penalty for reserving excess entangled pairs. Therefore, we can conclude that the over- and under-provision of entangled pairs can significantly impact the total cost.

\begin{figure*}[htb]
 \centering
 \captionsetup{justification=centering}
 \subfloat[The optimal total cost.]{\label{fig:optimal-solution-in-entangled-pair-resource-allocation}\includegraphics[width=0.33\textwidth]{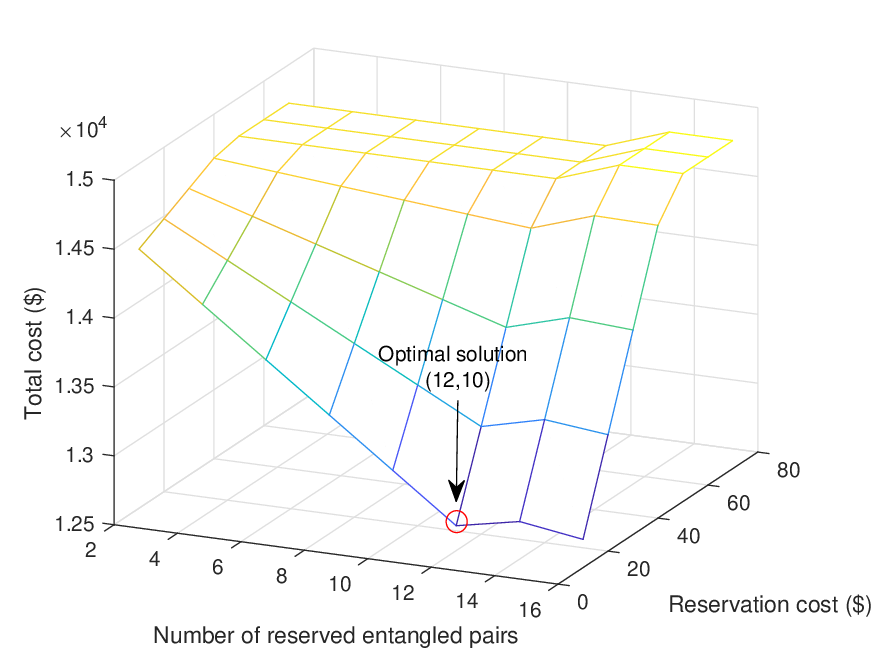}}
 \subfloat[Fidelity value comparison.]{\label{fig:fideity-requirement-entangled-pair}\includegraphics[width=0.33\textwidth]{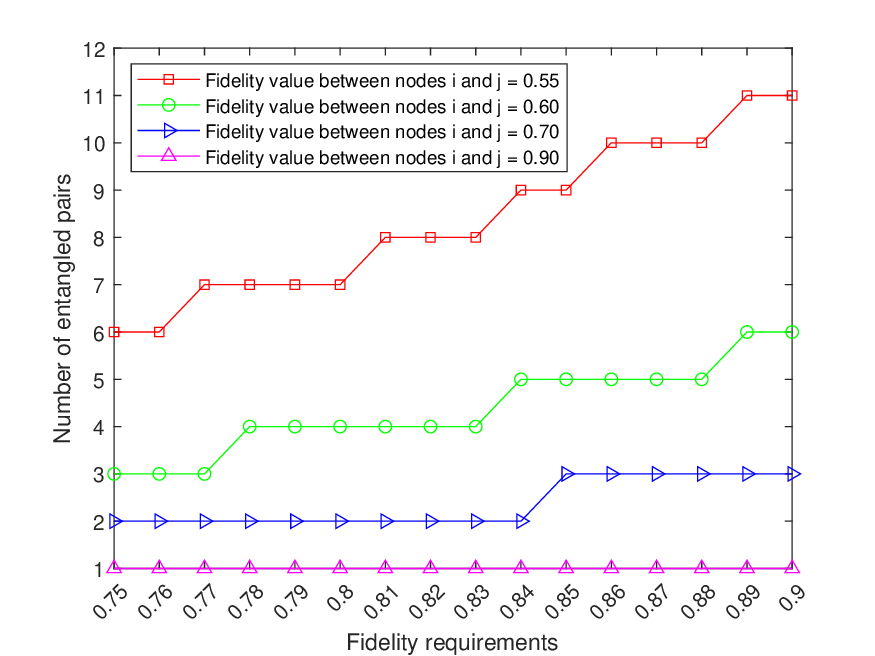}}
 \subfloat[The total costs.]{\label{fig:fidelity-demand-value}\includegraphics[width=0.33\textwidth]{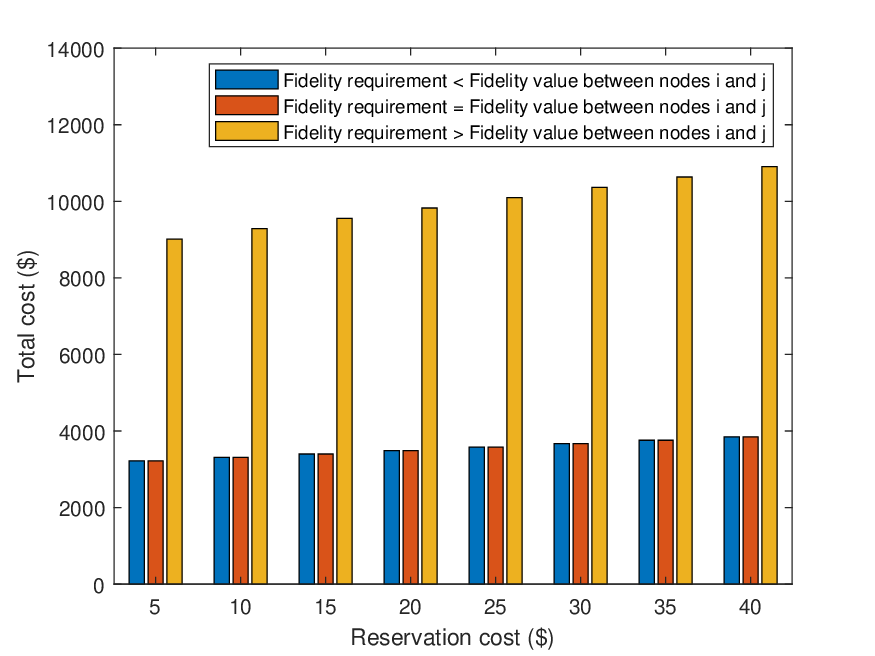}}
 \caption{(a) The optimal solution under different reservation costs and number of entangled pairs, (b) Comparison of entangled pair utilization of 4 fidelity values, and (c) Comparison of total cost of 3 fidelity requirements.}
 \label{fig:optimal-solution-entangled-pairs-total-costs}
 \vspace{-0.50cm}
\end{figure*} 

In addition, we investigate the performance of the proposed model in obtaining the optimal solution by varying the number of reserved entangled pairs and reservation costs. As shown in Fig.~\ref{fig:optimal-solution-entangled-pairs-total-costs}(a), the optimal cost can be attained when both the number of reserved entangled pairs and reservation cost increase, with the optimal solution achieved at 12 reserved entangled pairs and a reservation cost of \$10.

Figure~\ref{fig:optimal-solution-entangled-pairs-total-costs}(b) illustrates the performance of the proposed model in achieving the optimal number of entangled pairs for different fidelity requirements. The main observation is that the number of applied entangled pairs increases if the fidelity requirement exceeds the fidelity value between quantum nodes $i$ and $j$. This is due to the fact that higher fidelity requirements demand more entangled pairs to be used for entanglement purification in order to meet the required fidelity. Moreover, if the fidelity value between quantum nodes $i$ and $j$ is equal to or greater than the fidelity requirement, only one entangled pair is applied, which indicates that no entanglement purification is needed to enhance the fidelity value between those nodes.

We consider that, under different reservation costs, all fidelity values between quantum nodes $i$ and $j$ are 0.60 in the NSFNET topology to show the solution of the proposed model. As shown in Fig.~\ref{fig:optimal-solution-entangled-pairs-total-costs}(c), the total cost rises significantly when the reservation cost increases. Particularly, in the case of the fidelity requirement (FR) that is higher than the fidelity value (FV), the total cost in this case is higher than in the other two cases. This is because the number of entangled pairs is more utilized to perform the entanglement purification required to satisfy the fidelity requirement, resulting in a higher total cost. Therefore, from the aforementioned results shown in Figs.~\ref{fig:optimal-solution-entangled-pairs-total-costs}(b) and~\ref{fig:optimal-solution-entangled-pairs-total-costs}(c), we can conclude that fidelity values and fidelity requirements have a significant effect not only on the number of entangled pairs but also on the total cost.

\subsubsection{Qubit resource allocation} 

To demonstrate the varying execution times of quantum circuits with different numbers of qubits, we conduct experiments using Qiskit~\cite{quantum-fourier-transform-qiskit2022} to implement the QFT quantum circuits.

Figure~\ref{fig:qft-execution-time-optimal-solutions-total-costs}(a) demonstrates how the execution time of QDFT, implemented using Qiskit~\cite{quantum-fourier-transform-qiskit2022}, varies depending on the encoded numbers. As depicted in Fig.~\ref{fig:qft-execution-time-optimal-solutions-total-costs}(a), the execution time of QDFT is highest for an encoded number of 16383, due to the large number of qubits needed to represent and calculate the number in the transformation. Specifically, 14 qubits are required to represent the number 16383 in binary (i.e., 11111111111111) and the 14-qubit quantum circuit for the encoded number 16383 has a long depth. Thus, we can conclude that both the high encoded number and the long-depth quantum circuit have a direct impact on the execution time of the QDFT.

\begin{figure*}[htb]
 \centering
 \captionsetup{justification=centering}
 \subfloat[QDFT execution times.]{\label{fig:exetime-quantum-fourier-transform}\includegraphics[width=0.25\textwidth]{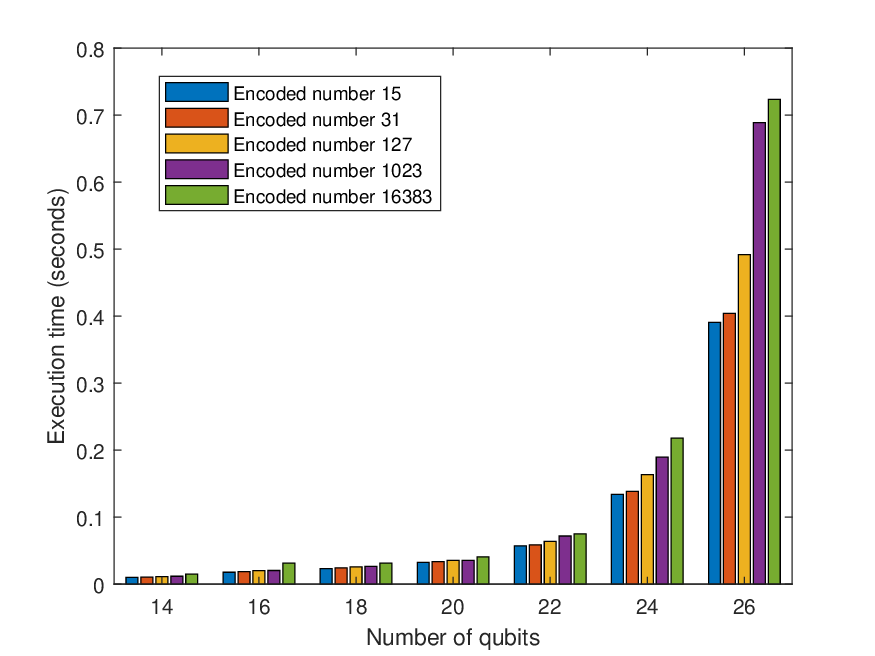}}
 \subfloat[The total and two-stage costs.]{\label{fig:optimal-solution-different-reserved-qubits}\includegraphics[width=0.25\textwidth]{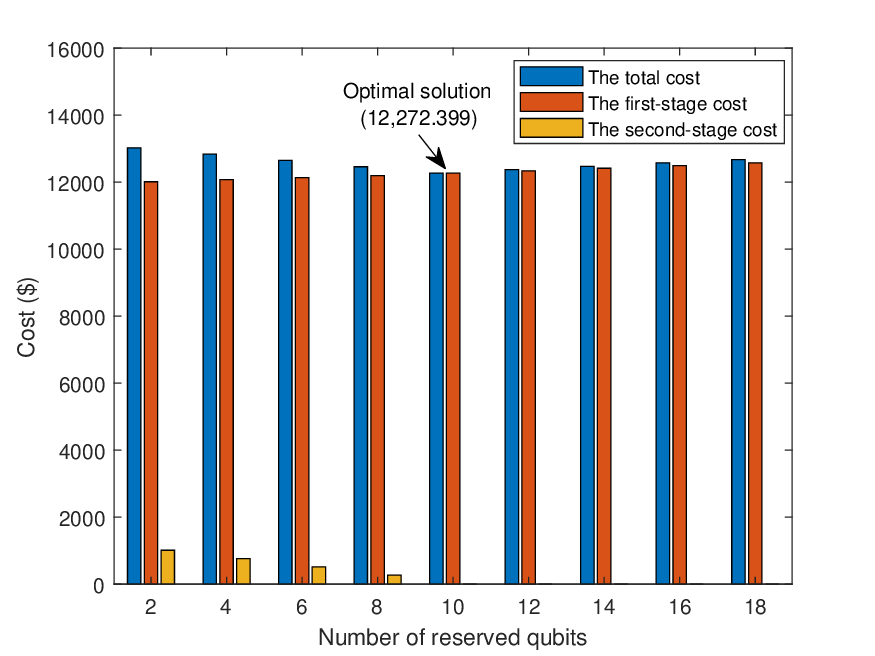}}
 \subfloat[The minimum total cost.]{\label{fig:optimal-solution-varying-qubits-waiting-time}\includegraphics[width=0.25\textwidth]{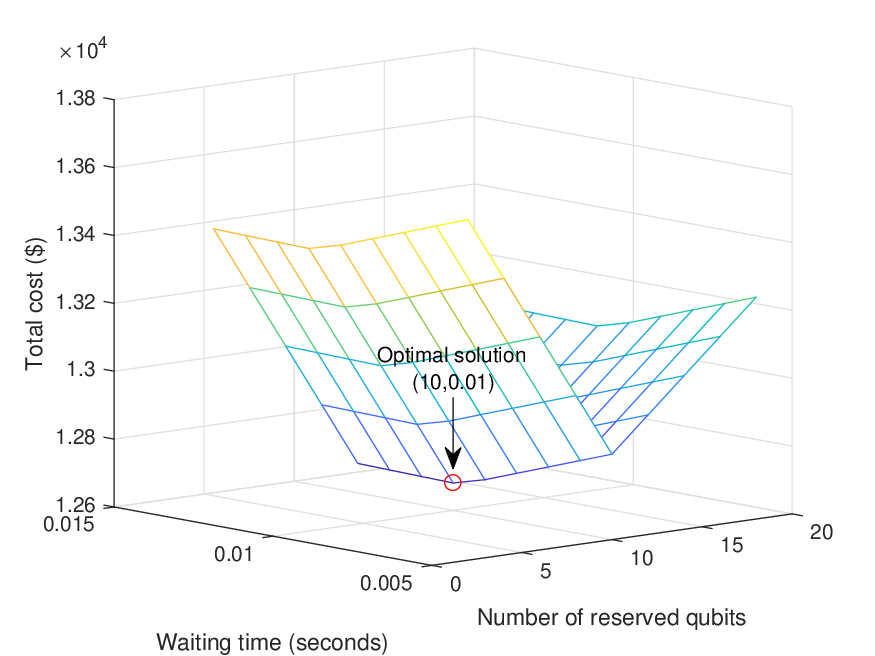}}
 \subfloat[The total costs.]{\label{fig:vary-reservation-cost-qubit}\includegraphics[width=0.25 \textwidth]{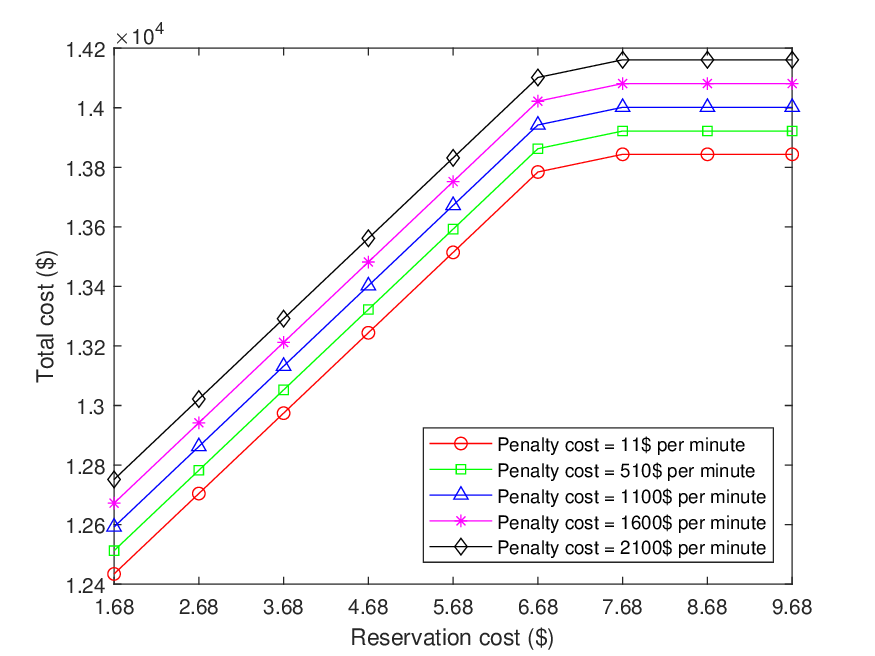}}
 \caption{(a) Comparison of QDFT's execution times with different encoded numbers, (b) The optimal solution in SP model, (c) The optimal solution under various arranged waiting times and number of reserved qubits, and (d) Comparison of total costs of 5 penalty costs.}
 \label{fig:qft-execution-time-optimal-solutions-total-costs}
 \vspace{-0.50cm}
\end{figure*} 

Figure~\ref{fig:qft-execution-time-optimal-solutions-total-costs}(b) illustrates that the cost in the first stage increases slightly as the number of reserved qubits increases, whereas the cost in the second stage gradually declines when the number of required qubits is observed in this stage. This is because the reservation qubit cost in the first stage is lower than the on-demand qubit cost in the second stage. Thus, to minimize costs, the qubit utilization in the second stage can be minimized while the qubit utilization in the first stage can be maximized. At the optimal solution with 10 reserved qubits, the first-stage cost and the total cost are \$12,272.399, but the second-stage cost becomes \$0. This is because the reserved qubit utilization meets the qubit demands while the on-demand qubit utilization remains unused. After reserving 12 qubits, both the total cost and the first-stage cost continue to rise slightly due to the penalty cost incurred for reserving more qubits than necessary. Thus, under-provisioning and over-provisioning of qubits have a notable impact on both the total cost and the first-stage cost. Figure \ref{fig:qft-execution-time-optimal-solutions-total-costs}(c) shows the optimal solution when the numbers of reserved qubits and arranged waiting times required by quantum circuits are varied. The optimal solution in Fig. \ref{fig:qft-execution-time-optimal-solutions-total-costs}(c) is achieved by the SP model, which consists of 10 reserved qubits and 0.01 seconds of waiting time, i.e., (10, 0.01).


We consider a scenario where the time it takes for a quantum computer from a provider to execute the QDFT circuit program ($E^{\mathrm{exe}}_{c,\rho,m,r}$) is longer than the time it takes for the program to wait ($\alpha_{r,c,\psi}$). To examine the impact on total cost, we explore the cost of qubits during the reservation phase and the cost of additional waiting time for quantum circuits. Figure~\ref{fig:qft-execution-time-optimal-solutions-total-costs}(d) shows the total costs at different penalty costs of over-waiting time. The graph shows that the total cost increases rapidly as the reservation cost goes up until it reaches \$6.68. This is because two factors contribute to the total cost: the cost of qubits during the reservation phase, and the cost of over-waiting time for quantum circuits ($y^{\mathrm{owt}}_{c,\rho,m,r,\psi} P^{\mathrm{wt}}_{c,\rho}$). During the reservation phase, qubits are used to meet the qubit demands, as the cost of qubits during the reservation phase is the lowest. The cost of over-waiting time is charged when quantum circuits have to wait longer than expected. Once the reservation cost reaches \$6.68, the total cost remains constant regardless of any penalty costs due to no waiting time. This is because, at this point, the number of qubits used during the on-demand phase is sufficient to meet the qubit demand, and the cost of using qubits during this phase is less than the cost of using qubits during the reservation phase. Therefore, raising the reservation cost has no effect on the total costs.

\begin{figure*}[htb]
 \centering
 \captionsetup{justification=centering}
 \subfloat[The qubit utilization.]{\label{fig:exe-qft}\includegraphics[width=0.25\textwidth]{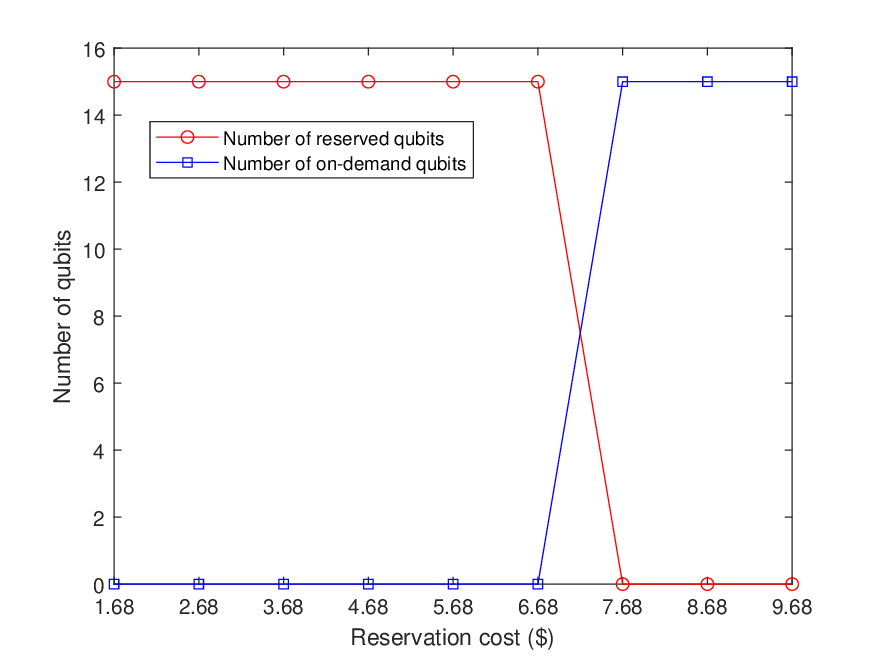}}
 \subfloat[Quantum circuit comparison.]{\label{fig:vary-waiting-time}\includegraphics[width=0.25\textwidth]{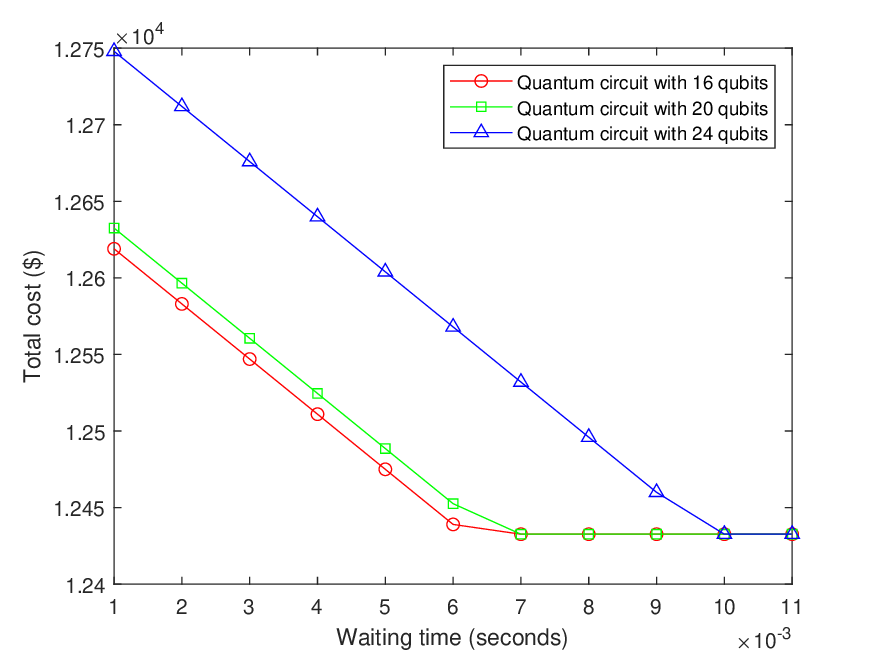}}
 \subfloat[The convergence.]{\label{fig:convergence-point-benders-decomposition}\includegraphics[width=0.25\textwidth]{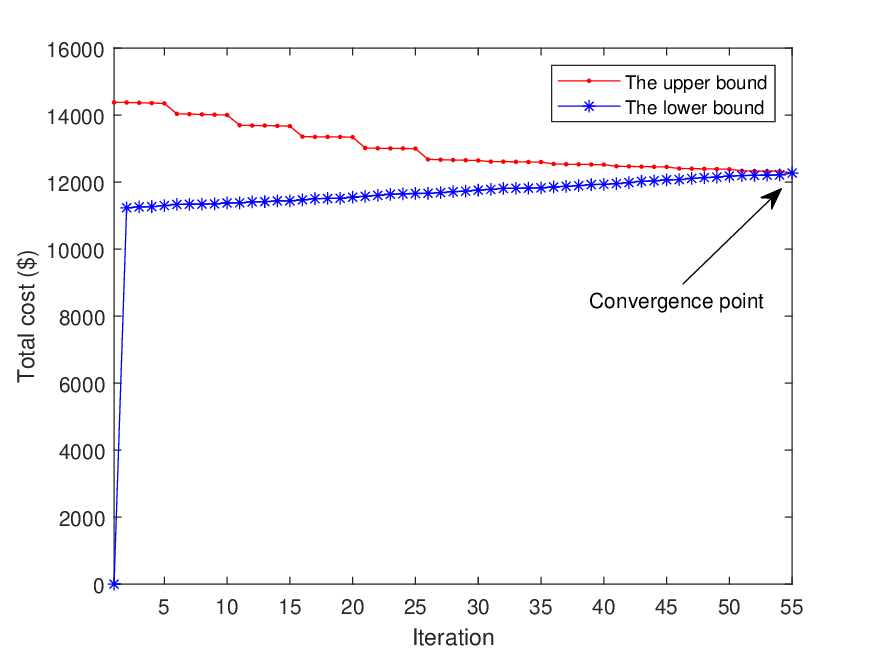}}
 \subfloat[The performance comparison.]{\label{fig:cost-comparison-three-models}\includegraphics[width=0.25 \textwidth]{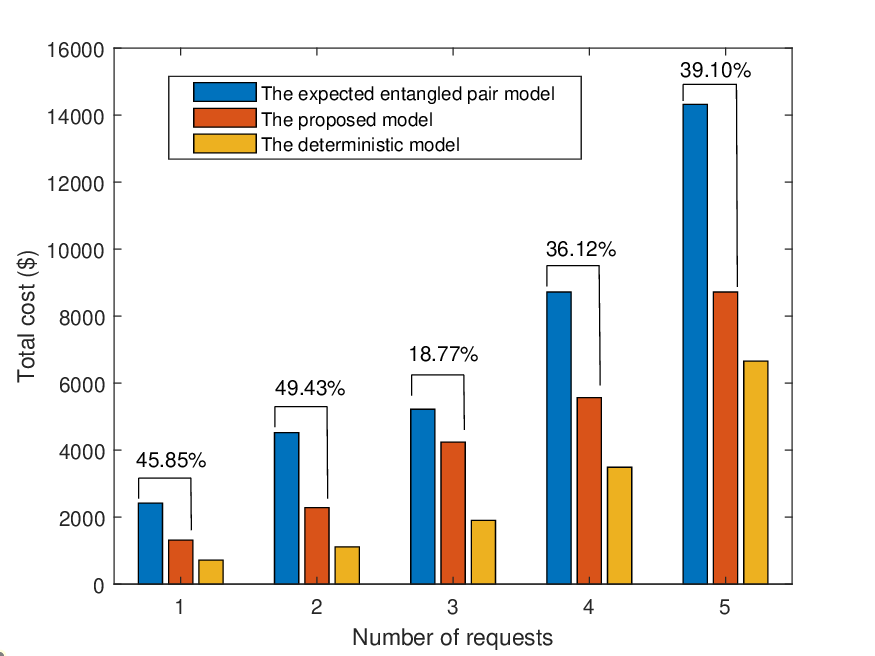}}
 \caption{(a) Comparison of qubit utilization in the reservation and on-demand phases, (b) Comparison of total costs of quantum circuits with different qubits, (c) The convergence of the upper and lower bounds by applying Benders decomposition algorithm, and (d) The performance comparison between the proposed model with two other models.}
 \label{fig:qubit-utilization-total-cost-convergence-cost-comparison}
 \vspace{-0.50cm}
\end{figure*} 

Based on the findings presented in Fig.~\ref{fig:qft-execution-time-optimal-solutions-total-costs}(d), Fig.~\ref{fig:qubit-utilization-total-cost-convergence-cost-comparison}(a) shows the number of qubits used during the reservation and on-demand phases. As expected, the qubits in reservation phase are utilized when the reservation cost falls between \$1.68 and \$6.68. However, when the reservation cost increases to \$7.68, the qubits in on-demand phase are used instead of the reserved ones and all total costs become constant.

To determine the total cost of the QDFT, we analyze the impact of the waiting time of quantum circuits. Specifically, we manipulate the waiting time of these circuits while keeping the encoded number 31 (which is represented as 11111 in binary) fixed. Figure~\ref{fig:qubit-utilization-total-cost-convergence-cost-comparison}(b) illustrates the total costs of running the QFT circuits under varying waiting times. The figure clearly shows that the 24-qubit quantum circuit has the highest total cost, while the 16-qubit quantum circuit has the smallest total cost, as the waiting time of the quantum circuits raises. Notably, for waiting times between 0.001 and 0.006 seconds, the total costs of all the circuits decrease significantly with an increase in the waiting time. This is because the penalty cost incurred due to the extra waiting time (i.e., $y^{\mathrm{owt}}_{c,\rho,m,r,\psi} P^{\mathrm{wt}}_{c,\rho}$) declines since the quantum computer of the provider is capable of completing quantum circuits before the time that quantum circuits require. In addition, the total costs become stable when it does not have additional waiting time for the quantum circuits. Specifically, the total costs of 16-qubit and 20-qubit quantum circuits are stable at waiting times of 0.007 seconds while the total cost of 24-qubit quantum circuit is stable at 0.01 seconds. Thus, we can conclude that the waiting time of quantum circuits is a critical factor affecting the total costs, as indicated in Fig.~\ref{fig:qubit-utilization-total-cost-convergence-cost-comparison}(b).

\subsubsection{Benders decomposition and cost comparison}

Figure~\ref{fig:qubit-utilization-total-cost-convergence-cost-comparison}(c) illustrates the convergence of the bounds obtained through the Benders decomposition algorithm. In each iteration, the lower and upper bounds are adjusted. The algorithm converges at iteration 55. The optimal solution obtained from the Benders decomposition algorithm is the same as the one obtained by solving the SP model without decomposition. We observe that while the subproblems can be solved efficiently due to their smaller number of variables and parallelization, the master problem requires a substantial amount of time as more Benders cuts need to be added.

We compare the performance of the proposed model with two models: the expected entangled pair model and the deterministic model. For the expected entangle pair model, we consider the fidelity demands as expected values ($\bar{\delta}_{r,c}$) and solve the model in Eqs.~(\ref{eq:def_obj})-(\ref{eq:def_const15}) using these expected values. For the deterministic model, we consider the fidelity demands as exact values ($\hat{\delta_{r,c}}$) and solve the model in Eqs.~(\ref{eq:def_obj})-(\ref{eq:def_const15}) using these exact values. From Fig.~\ref{fig:qubit-utilization-total-cost-convergence-cost-comparison}(d), it is clear that the proposed model achieves the minimum cost compared to the expected entangled pair model as the number of requests increases. In particular, the proposed model can decrease the total cost by 45.85\%, 49.43\%, 18.77\%, 36.12\%, and 39.10\% for the number of requests 1, 2, 3, 4, and 5, respectively. This demonstrates the significant cost savings that can be achieved by using the proposed model. Nevertheless, the proposed model performs worse than the deterministic model since the latter uses exact fidelity demands to achieve the solution, while the former uses statistical fidelity demands. Nevertheless, the proposed model is more practical than the deterministic model since, in reality, it is difficult to know the exact entangled pair demands, which are necessary inputs for the deterministic model to yield the solution.

\section{Conclusion}
\label{sec:conclusion}
In QCC, users will require quantum computing resources as well as entanglement routing with a fidelity guarantee for the communication between users and providers. Therefore, the quantum resource operator, implemented by the proposed model, can provide quantum computing resources and fidelity-guaranteed entanglement routing, jointly optimizing them to minimize the overall cost. We have proposed the joint entangled pair and qubit resource allocation, and entanglement routing with a fidelity guarantee in the QCC. Our model is to provision entangled pairs and fidelity-guaranteed entanglement routing to the quantum network, while qubit resources are provisioned to QCC providers. We have formulated the resource allocation based on the two-stage SP model for entangled pairs, fidelity-guaranteed entanglement routing, qubit resources for quantum circuits, and the minimum waiting time of quantum circuits, which considers uncertainties of fidelity requirements, qubit requirements, and circuit waiting time to achieve the optimal total cost. In addition, we have applied Benders decomposition algorithm to break down the resource allocation model into sub-models that are solved simultaneously. The experimental results have indicated that the proposed model achieves the optimal total cost in terms of entangled pairs, entanglement routing, qubits, and minimum circuit waiting time, surpassing the expected entangled pair model by at least 18.77\%.

\end{document}